\documentclass[aps,pra,twocolumn,groupedaddress,floatfix]{revtex4-1}

\usepackage{graphicx}
\usepackage{dcolumn}
\usepackage{bm}
\usepackage{amsmath}
\usepackage{xcolor}

\newcommand{\fdlu}{$\text{f}\!\downarrow\!\ell\!\uparrow\,$}
\newcommand{\fuld}{$\text{f}\!\uparrow\!\ell\!\downarrow\,$}
\newcommand{\fulu}{$\text{f}\!\uparrow\!\ell\!\uparrow\,$}

\begin{document}

\title{
\textcolor{black}{
Two-point momentum correlations of few ultracold quasi-one-dimensional trapped fermions:
Diffraction patterns}
}

\author{Benedikt B. Brandt}
\email{benbra@gatech.edu}
\author{Constantine Yannouleas}
\email{Constantine.Yannouleas@physics.gatech.edu}
\author{Uzi Landman}
\email{Uzi.Landman@physics.gatech.edu}

\affiliation{School of Physics, Georgia Institute of Technology,
             Atlanta, Georgia 30332-0430, USA}

\date{21 August 2017; PRA {\bf 96}, 053632 (2017)}

\begin{abstract}
Spatial and momentum correlations are important in the analysis of the quantum states and different phases 
of trapped ultracold atom systems as a function of the strength of interatomic interactions. Identification 
and understanding of spin-resolved patterns exhibited in two-point correlations, accessible directly by 
experiments, are key for uncovering the symmetry and structure of the many-body wavefunctions of the trapped 
system. Using the 
\textcolor{black}{full} 
configuration interaction method for exact diagonalization of the many-body Hamiltonian of $N=2-4$ fermionic 
atoms trapped in single, double, triple, and quadruple wells, we analyze both two-point momentum and space 
correlations, as well as associated noise distributions, for a broad range of interparticle contact 
repulsion strengths and interwell separations, unveiling characteristics allowing 
\textcolor{black}{
insights into the transition, via an intermediate phase,  from the non-interacting Bose-Einstein condensate to 
the weakly interacting quasi-Bose-Einstein regime, and from the latter to the strong-repulsion} 
Tonks-Girardeau (TG) one. The {\it ab-initio} numerical predictions are shown to agree well with the results of
a constructed analytical model employing localized displaced Gaussian functions to represent the $N$ fermions.
The two-point momentum correlations are found to exhibit damped oscillatory diffraction behavior. This 
diffraction behavior develops fully for atoms trapped in a single well with strong interatomic repulsion in the 
TG regime, or for atoms in well-separated multi-well traps.
\textcolor{black}{
Additionally, the two-body momentum correlation and noise distributions are found to exhibit 
``shortsightedness'', with the main contribution coming from nearest-negihboring particles.}
\end{abstract}

\maketitle

\section{Introduction}

Recent groundbreaking experimental progress in time-of-flight measurements is providing an abundance
of information for the two-point and higher-order momentum correlations of one-dimensional systems with a
large number of trapped bosons \cite{altm04,new1,new2,bloc05,bloc06,spie07,bouc12,new4,new5,new6,bouc16,hodg17}. 
Such information reflects directly the nature of the 
correlated many-body wave function and can be used as a tool to probe theoretical models and methodologies; 
\textcolor{black}{
e.g., it has been found \cite{bouc16} that 1D boson systems deviate from the predictions \cite{altm09} (see
also Ref.\ \cite{deml09}) of the Bogoliubov theory \cite{bogo47} in the quasi Bose-Einstein condensate 
(QBEC) regime in-between the ideal-Bose gas and the strongly correlated Tonks-Girardeau (TG) regimes.}

Motivated by the above developments and the experimental advances in controlling a few deterministically 
prepared fermions \cite{joch11,joch15}, we present exact configuration-interaction (CI) results for the 
two-point momentum, as well as spatial, correlations of a few ultracold fermionic 
atoms confined in quasi-1D single- and multi-well traps. 
\textcolor{black}{
Theoretical investigations of two-point space correlations for a few fermions (electrons) confined in 
semiconductor quantum dots abound; for a small sample of earlier literature, see Refs.\ \cite{yann07,li07,li09}. 
Several studies of two-point space correlations have also been reported for a few trapped ultracold atoms 
\cite{yann04,yann07.2,garc14,yann15,yann16,poll17},}  
but the corresponding theoretical predictions for the momentum correlations, which can be directly compared to 
time-of-flight measurements, are still missing. (Studies of
two-point momentum correlations for bosons in the TG regime are also lacking \cite{bouc16}.) 
 
Based on configuration-interaction (CI) calculations, this paper provides a complete range of 
{\it ab-initio\/} two-point momentum-correlation maps (including noise distributions) for a small number $N$ 
of ultracold trapped fermions, as a function of the strength $g$ of the short-range repulsion, the total spin 
($S$, $S_z$), and for both the cases of a single-well or a multi-well trap with different interwell
separations. One of our main findings shows that at the Tonks-Girardeau 
regime the momentum correlations exhibit a signature pattern of damped diffraction 
(interference) oscillations associated with a typical distance-scale arising 
from the emergent spatial particle localization in this regime. Control of the typical diffraction length is 
achieved by increasing the interwell separation in singly-occupied multi-well traps, resulting in a larger 
number of visible diffraction minima. The diffraction behavior of two-point momentum correlations
was reported early on by Coulson for the case of the natural H$_2$ molecule, aiming at gaining 
momentum-space insights into molecular bonding \cite{coul41}. It readily lends an interpretation of the 
Tonks-Girardeau regime as a special limit in the context of a
general unified theory of Wigner-molecule formation in finite systems with strongly repulsive interparticle 
interactions \cite{yann07,yann04,yann07.2,yann15,yann16}, in particular, here, ultra-cold Wigner molecules 
(UCWM) \cite{yann15,yann16}.

The plan of the paper is as follows: We begin in Sec.\ \ref{meth} with a short description of the theoretical 
methodology developed and used in this work, including: (i) the CI method for exact diagonalization of the 
many-body Hamiltonian of $N$ optically-trapped ultracold atoms (Sec.\ \ref{hci}), (ii) {\it ab-initio\/} 
numerical calculations of one- and two-point real-space and momentum-space correlation, and noise, functions, 
(Sec.\ \ref{defs}) and (iii) analytic modeling of the above-noted correlation functions, illustrated in detail 
for the case of two-particles with a discussion of two-particle interference pattern and correlation-map 
derivation (Sec.\ \ref{anmdtp}). In Sec.\ \ref{res} we display and discuss 
the results of our CI calculations for the following cases: (A)  Two fermions in a single quasi-1D well with an 
emphasis on the dependence on the inter-atomic interaction strength and the Tonks-Girardeau limit, 
\textcolor{black}{
including an illustration of the shortsightedness of the two-body momentum noise distribution,}
(B) Two fermions in a quasi-1D double-well, and (C) Three fermions trapped in quasi-1D single or triple wells, 
with an emphasis on spin-resolved, versus spin-unresolved, two-point correlation maps. Sec.\ \ref{comp} is 
devoted to comparisons between the {\it ab-initio\/} CI numerically calculated two-point spin-resolved (and spin 
unresolved) correlations in real and momentum space for two, three, and four ultracold fermionic atoms trapped 
in double, triple, and quadruple wells. We summarize our wok in Sec.\ \ref{conc}. In the Appendices, we give 
explicit expressions for the analytically-derived two-point correlation functions for two (including also 
two-point noise functions), three, and four atoms, as well as the effective Heisenberg Hamiltonians for three 
and four well-localized atoms. 
\textcolor{black}{
We note that, for a small number of repelling trapped particles (electrons in
semiconductor quantum dots and ultracold fermions or bosons), the mapping of the microscopic many-body 
Hamiltonian onto spin-chain-type, effective Heisenberg Hamiltonians has been demonstrated recently and
it constitutes an ongoing active area of research; for electrons in semiconductor quantum dots 
see Refs.\ \cite{li07,li09}, for  ultracold fermions or bosons in quasi-1D traps see Refs.\ 
\cite{yann15,yann16,zinn14,deur14,zinn15,bruu15,mass15,murm15,murm15.2,pu15,cui16}.  
}

\section{Methods}
\label{meth}

\subsection{Many-body Hamiltonian}
\label{hci}

In this paper we employ the configuration-interaction (CI) method (referred to also as exact diagonalization 
method) to determine the solution of the two-dimensional $N$-body fermionic Hamiltonian

\begin{align}
H_{\rm MB}=\sum_{i=1}^N H(i)+\sum_{i=1}^N\sum_{j>i}^N V({\bf r}_i,{\bf r}_j),
\label{mbh}
\end{align}

\noindent where $H(i)$ represents the single particle part of the many-body Hamiltonian and 
$V({\bf r}_i,{\bf r}_j)$ represents the interaction term, with ${\bf r}_i \equiv (x_i,y_i)$ 
and ${\bf r}_j \equiv (x_j,y_j)$ being the space 
coordinates of the $i$th and $j$th particle respectively. The single particle part $H(i)$ of the Hamiltonian 
contains the kinetic energy term and a single-particle external confining potential; in this paper we consider 
double-, triple-, and quadruple-well confinements in a linear arrangement. 

The external confining potential has been extensively described in \cite{yann15,yann16}. 
The relevant potential parameters for this paper are the inter-well spacing $d_w$,
which is indicated in our figures (obviously, $d_w=0$ for a single 
well) and the value of $\epsilon_b$ (determining the interwell barrier height) which is taken to be $0.5$ 
throughout the paper. Each of the parabolic confining wells is characterized by two harmonic frequencies, 
$\hbar\omega_x$ (along the long $x$-axis of the well) $<<\hbar\omega_y$ (along the $y$ direction), 
resulting in a (quasi-onedimensional) needle-like shape confinement, so that only the lowest-in-energy 
single-particle space orbital in the $y$-direction is populated. In our calculations 
$\hbar\omega=\hbar\omega_x =1$ kHz, and $\hbar\omega_y =100$ kHz (hereafter we drop for convenience the 
subscript $x$). 
\textcolor{black}{
In experimental realizations of quasi-1D (needle-shape) confinement, a similar strategy is employed, with a 
ratio of 10-250 between the transverse and longitudinal confining frequencies \cite{bouc16,joch11,joch15}.} 
The interaction term is given by 
\begin{align}
V({\bf r}_i,{\bf r}_j)=\frac{g}{\sigma^2 \pi}e^{-({\bf r}_i-{\bf r}_j)^2/\sigma^2}.
\label{tbi}
\end{align}
\noindent In this paper we use $\sigma=\sqrt{2}l_0/10=0.1833$ $\mu$m where $l_0$ is the harmonic 
oscillator length $l_0^2=\hbar/(M_{^6{\rm Li}}\omega)$ with $M_{^6{\rm Li}}=10964.90m_e$ being the mass 
of $^6{\rm Li}$. The division of $l_0$ in the expression for $\sigma$ by a factor of 
ten is motivated by the need to model short-range, contact-like interactions. Any Gaussian width $\sigma$ 
that is much smaller than the harmonic oscillator length $l_0$ along the $x$-direction is suitable and yields 
essentially identical final results. 

Common values for $g$ in this paper are given below, in both atomic units (energy in Rydberg and length in 
Bohr-radius units, $a_0$) and in $\hbar\omega\; l_0^2$ (often used in describing experimental setups):

\begin{center}
\begin{tabular}{ c|c } 
 $g\; [\text{Ry}\;a_0^2]$ & $g\; [\hbar\omega\; l_0^2]$ \\ 
 \hline
 0.0001 & 0.5486 \\ 
 0.001 & 5.486 \\ 
 0.01 & 54.86 
\end{tabular}
\end{center}

\subsection{Configuration-interaction method, correlation functions, and noise distributions}
\label{defs}

Details of our CI methodology and the single-particle external confining potential can be found in Refs.\
\cite{yann15,yann16,yann07.2,yann07}. A CI many-body wave function $\Phi_{\rm CI}^N$ has good total-spin $S$ 
and spin-projection $S_z$ quantum numbers
and is specified as a superposition of Slater determinants $\Psi^N$ built out of spin-and-space orbitals 
$\varphi_i({\bf w})$ [${\bf w} \rightarrow ({\bf r},\sigma)$] belonging to a given single-particle basis set; 
i.e.,
\begin{align}
\Phi_{\rm CI}^N = \sum_J C_J \Psi^N_J.
\label{ciwf}
\end{align} 

\textcolor{black}{
In expansion (\ref{ciwf}), we use all the determinants that can be built out from a basis set of $K$ 
single-particle spin-orbitals. The number $K$ is allowed to increase stepwise. When the result converges with 
respect to the number ($K$) of the spin-orbitals in the basis, one obtains an exact diagonalization of the 
many-body Hamiltonian defined in Eq.\ (\ref{mbh}) \cite{szosbook,paun67}; the converged CI is often termed 
``full CI''.}

Given an $N$-particle wavefunction $\Phi({\bf w}_1,{\bf w}_2,\dots,{\bf w}_N)$, the two-point 
real-space correlation function normalized to unity is given as 

\begin{align}
\begin{split}
\mathcal{P}({\bf w}_1,{\bf w}_1',{\bf w}_2,{\bf w}_2') = 
&\int_{-\infty}^\infty\Phi^\dagger({\bf w}_1',{\bf w}_2',{\bf w}_3,\ldots,{\bf w}_N) \\
& \times \Phi({\bf w}_1,{\bf w}_2,{\bf w}_3,\ldots,{\bf w}_N)d{\bf w}_3\ldots d{\bf w}_N,
\label{eq:corr_function_def_2}
\end{split}
\end{align}

\noindent
where ${\bf w}_i$ represents the space ${\bf r}_i$ and spin coordinate $\sigma_i$ of particle $i$. The 
one-point real-space correlation function normalized to unity is obtained as

\begin{align}
\begin{split}
\rho({\bf w}_1,{\bf w}_1')& = \int_{-\infty}^\infty
\Phi^\dagger({\bf w}_1',{\bf w}_2,{\bf w}_3,\ldots,{\bf w}_N)\\
& \;\;\;\; \times \Phi({\bf w}_1,{\bf w}_2,{\bf w}_3,\ldots,{\bf w}_N)
d{\bf w}_2\ldots d{\bf w}_N\\
& =\int_{-\infty}^\infty\mathcal{P}({\bf w}_1,{\bf w}_1',{\bf w}_2,{\bf w}_2)d{\bf w}_2.
\label{eq:corr_function_def}
\end{split}
\end{align}

We note that the physically relevant quantities for the purpose of this paper are the diagonal parts of the 
correlation functions. The off-diagonal parts are used as auxiliary quantities to Fourier transform from 
real-space to momentum space and vice versa. For the two-body and one-body momentum correlation functions, the 
physically relevant diagonal parts are obtained via the following Fourier transforms:

\begin{align}
\begin{split}
{\cal G}  ({\bf q}_1,{\bf q}_2) = & \frac{1}{4\pi^2}
\int e^{ -i {\bf q}_1 \cdot ( {\bf w}_1-{\bf w}_1') } e^{-i {\bf q}_2 
\cdot ( {\bf w}_2-{\bf w}_2')} \\
& \times  {\cal P} ({\bf w}_1, {\bf w}_1', {\bf w}_2, {\bf w}_2')  
d{\bf w}_1 d{\bf w}_1' d{\bf w}_2 d{\bf w}_2',
\label{tbmc}
\end{split}
\end{align}
and
\begin{align}
\tau({\bf q})=\frac{1}{2\pi} \int_{-\infty}^\infty e^{ -i {\bf q} \cdot ({\bf w}_1-{\bf w}_1') }
\;\rho({\bf w}_1,{\bf w}_1')d{\bf w}_1 d{\bf w}_1',
\label{obmc}
\end{align}

\noindent
where ${\bf q}_i$ represents the momentum ${\bf k}_i$ and spin coordinate $\sigma_i$ of particle $i$. Once one 
has obtained the one-point and two-point correlation functions, the calculations of noise distributions in real 
${\mathcal{P}_\mathcal{N}}$ and momentum $\mathcal{G}_\mathcal{N}$ space are straightforward:

\begin{align}
\mathcal{P}_\mathcal{N}({\bf w}_1,{\bf w}_2)&=\mathcal{P}({\bf w}_1,{\bf w}_1,{\bf w}_2,{\bf w}_2)-
\rho({\bf w}_1,{\bf w}_1)\rho({\bf w}_2,{\bf w}_2),
\label{pn}
\end{align}
\noindent and
\begin{align}
\mathcal{G}_\mathcal{N}({\bf q}_1,{\bf q}_2)&=\mathcal{G}({\bf q}_1,{\bf q}_2)-
\tau({\bf q}_1)\tau({\bf q}_2).
\label{gn}
\end{align}

\subsection{Analytic modeling: Two-particle interference pattern and correlation map derivation}
\label{anmdtp}

The microscopic numerical CI evaluation of the correlation functions defined in Sec.\ \ref{defs} are 
complemented by analytical expressions extracted from a simple model of localized particles represented by 
displaced Gaussian orbitals in the spatial Hilbert space. 
In this section and the Appendices, we display such analytical modeling for 
two, three, and four fermions. Here we illustrate in some detail the derivation of such interference 
formulas for two particles, allowing a rather immediate generalization to more complex cases, like $N=3$ and 
$N=4$ particles;  the analytical expressions for the noise function for $N=2$ are given in Appendix \ref{nm2p}, 
and the two-point real-space and momentum-space correlation functions for $N=3$ and $N=4$ particles, as well as 
the corresponding Heisenberg model Hamiltonians, are given in Appendices \ref{sr34}, \ref{su34}, and 
\ref{sec:hsberg_hamil_and_eigenfunction}. For simplicity the calculations are done here in one-dimension, with 
the generalization to higher dimensions being rather straightforward.

\textcolor{black}{
This analytic modeling grasps the main physics of particle localization in the case of repulsive two-body
interaction. Moreover, it offers immediate insight why the particle localization (induced by the separated 
wells, as well as by Wigner-molecule formation in the case of a single well) produces a characteristic signature
of a damped diffraction pattern in the two-point momentum correlations. In this modeling, we assume that
the spatial part of the $j$th particle is approximated by a displaced Gaussian function (each localized at a 
position $d_j$),
\begin{align}
\psi_j(x)=
\frac{1}{(2 \pi)^{1/4}\sqrt{s}}\exp \left( - \frac{(x-d_j)^2}{4s^2} \right),
\label{disorb}
\end{align}
\noindent where $s$ denotes the width of the Gaussian functions. 
}
 
\textcolor{black}{
The single-particle orbital $\psi_j(k)$ in the momentum Hilbert space is given by the Fourier transform
of $\psi_j(x)$, namely $\psi_j(k)=(1/\sqrt{2\pi}) \int_{-\infty}^\infty \psi_j(x) \exp(ikx) dx$. 
Performing this Fourier transform, one finds
\begin{align}
\psi_j(k)=
\frac{2^{1/4}\sqrt{s}}{\pi^{1/4}}\exp ( - k^2 s^2 ) \exp(i d_j k).
\label{disorbk}
\end{align}
\noindent
Eq.\ (\ref{disorbk}) explicitly illustrates how the displacement $d_j$ in the real space (associated with 
particle localization) generates a plane-wave behavior [the factor $\exp(i d_j k)$] in the momentum
space. As is calculated explicitly below, in the case of several localized particles, these plane-wave
factors produce interference diffraction patterns in the two-body momentum correlations that depend in general 
on the characteristic mutual distances $2d_{ij}=d_i-d_j$ between the particles. One of the main conclusions from
the analytic modeling, however, is that these interference patterns are primarily controlled by the minimum 
distance $2d=d_1-d_2$ between adjacent particles. Moreover the interference patterns do not extend in the
full range of momenta $-\infty < k < \infty$, because they are damped by the damping factor ${\cal A}(k) = 
\exp( - 2 k^2 s^2 ) $ (see below) which is the square of the exponential in Eq.\ (\ref{disorbk}).
}

\textcolor{black}{
As a consequence, the two-point momentum correlation function (derived from the many-body wavefunction) focuses 
on properties associated with the smallest interparticle distance in the multiparticle system -- that is, it 
provides information associated with nearest-neighbor particles.} 
\textcolor{black}{This shortsightedness}
\textcolor{black}{
suggests that information 
extracted from investigations of two-point momentum distributions for finite (small) numbers of particles (for 
which reliable many-body results can be obtained computationally, i.e., using full CI and exact Hamiltonian 
diagonalization as described in this work), could enhance in a significant way the  understanding of properties 
of larger systems under similar conditions (for example, similar interparticle interaction strength) for which 
reliable many-body solutions are complex and often unknown (see below).
}

\textcolor{black}{
To compare the spin-resolved two-body CI correlations with those derived from the analytic model, we need
to guarantee that the approximate many-body wave functions of the analytic model conserved the total spin
$S$ and its projection $S_z$, a property that is automatically satisfied in the microscopic CI approach in
the absence of energy degeneracies. In the analytic modeling, we need to construct appropriate total-spin
eigenfunctions which obey the branching diagram \cite{li09,yann16,paun00} of total-spin multiplicities and 
other properties described in detail in Ref.\ \cite{paun00}.} 
\textcolor{black}{
For localized particles, where (as described above) the spatial part of the wave function can be approximated 
by the displaced Gaussian functions, the complex task of determining the appropriate total-spin components 
simplifies because these can be readily obtained as total-spin eigenfunctions through the exact eigenvector 
solutions of an effective Heisenberg Hamiltonian 
\cite{li07,li09,yann15,yann16,zinn14,deur14,zinn15,bruu15,mass15,murm15,murm15.2,pu15,cui16}.
We stress that we need to obtain here exact solutions of the Heisenberg Hamiltonian, a task that is feasible
for a small number $N$ of particles. It is pertinent also to remark explicitly that, for the purpose of this 
work, specifically for analyzing the properties of the two-body correlation functions (particularly in the 
strongly-interacting highly-correlated  regime), use of the most familiar mean-field solutions \cite{auer94}, 
most often employed for the description of larger particle systems, will not suffice. 
}

As mentioned earlier, we address in this section the case of two ($N=2$) localized fermions, for which
the corresponding effective Heisenberg Hamiltonian is very simple, i.e., 

\begin{align}
H=J\;{\bf S_1}\cdot{\bf S_2}-\frac{J}{4},
\end{align}

\noindent where ${\bf S_1}$ and ${\bf S_2}$ are spin operators and $J$ is the coupling constant. Using the spin-primitives $|\uparrow\downarrow\rangle$ and $|\downarrow\uparrow\rangle$ this Hamiltonian can be expressed in matrix form as

\begin{align}
H=J\left(
\begin{array}{cc}
 0 & \frac{1}{2} \\
 \frac{1}{2} & 0 \\
\end{array}
\right),
\end{align}

\noindent with eigenvalues $e_1,e_2$ and eigenvectors $v_1,v_2$

\begin{align}
e_1&=-J/2,\\
e_2&=J/2,\\
v_1&=\frac{1}{\sqrt{2}}(|\uparrow\downarrow\rangle-|\downarrow\uparrow\rangle),
\label{v1}\\
v_2&=\frac{1}{\sqrt{2}}(|\uparrow\downarrow\rangle+|\downarrow\uparrow\rangle).
\end{align}

Naturally, as mentioned earlier, the above Heisenberg-model solutions pertain to the spin part of the 
wavefunctions. To include the spatial component of the wavefunctions we need to associate each 
spin primitive (i.e., $|\uparrow\downarrow\rangle$ or $|\uparrow\downarrow\rangle$) with a determinant of 
spin-orbitals $\psi_{j\sigma}(x)$ ($j$ denotes the $j$-th space orbital, $\sigma$ represents the spin). 
The corresponding determinants ${\cal D}$'s to each primitive are 

\begin{widetext}
\begin{align}
|\uparrow\downarrow\rangle &\longrightarrow \mathcal{D}_{\uparrow\downarrow}(x_1,x_2)=
\frac{1}{ \sqrt{2!} }\left|
\begin{array}{cc}
 \psi_{1\uparrow}(x_1) & \psi_{2\downarrow}( x_1) \\
 \psi_{1\uparrow}(x_2) & \psi_{2\downarrow}(x_2) \\
\end{array}
\right|=\frac{1}{ \sqrt{2!} }(\psi_{1\uparrow}(x_1) \psi_{2\downarrow}(x_2)-
\psi_{1\uparrow}(x_2)\psi_{2\downarrow}(x_1)) \label{d2_1}\\
|\downarrow\uparrow\rangle &\longrightarrow \mathcal{D}_{\downarrow\uparrow}(x_1,x_2)=
\frac{1}{\sqrt{2!}}\left|
\begin{array}{cc}
 \psi_{1\downarrow}(x_1) & \psi_{2\uparrow}(x_1) \\
 \psi_{1\downarrow}(x_2) & \psi_{2\uparrow}(x_2) \\
\end{array}
\right|=\frac{1}{\sqrt{2!}}(\psi_{1\downarrow}(x_1) \psi_{2\uparrow}(x_2)-
\psi_{1\downarrow}(x_2)\psi_{2\uparrow}(x_1)). \label{d2_2}
\end{align}
\end{widetext}

We can use the two determinants in Eqs.\ (\ref{d2_1}) and (\ref{d2_2}) together with the ground-state 
eigenvector $v_1$ [Eq.\ (\ref{v1})] to form the Heitler-London \cite{hl27,yann02,yann02.2} ground-state wave 
function $\Phi_{\rm HL}({\bf x}_1,{\bf x}_2)$ and the associated two-body correlation function [see Eq.\ 
(\ref{eq:corr_function_def_2})] $\mathcal{P}_{\rm HL} ({\bf x}_1,{\bf x}_1',{\bf x}_2,{\bf x}_2')$, where the 
boldfaced ${\bf x} \rightarrow (x,\sigma)$,

\begin{align}
& \Phi_{\rm HL}({\bf x}_1,{\bf x}_2) =\frac{1}{{\cal N}_2} 
\frac{1}{\sqrt{2}}(\mathcal{D}_{\uparrow\downarrow}(x_1,x_2)-
\mathcal{D}_{\downarrow\uparrow}(x_1,x_2)) \\
& \mathcal{P}_{\rm HL} ({\bf x}_1,{\bf x}_1',{\bf x}_2,{\bf x}_2') =
\Phi_{\rm HL}^\dagger({\bf x}_1',{\bf x}_2')\Phi_{\rm HL}({\bf x}_1,{\bf x}_2),
\end{align} 
\noindent
where the factor $1/{\cal N}_2$ normalizes the HL wave function. Specifically, ${\cal N}_2=\sqrt{1+S_{12}^2}$,
where $S_{12}$ is the overlap of the two (in general non-orthogonal) localized space orbitals;
see Eqs.\ (10) and (11) in Ref.\ \cite{yann02.2}. 

We stress here that for the case of more than two particles the additional particle coordinates need
to be integrated out to arrive at the two-point correlation function, e.g., for three particles: 

\begin{align}
\begin{split}
& \mathcal{P}({\bf x}_1,{\bf x}_1',{\bf x}_2,{\bf x}_2') \\
& = \int_{-\infty}^\infty\Phi_{\rm gs}^\dagger({\bf x}_1',{\bf x}_2',{\bf x}_3)
\Phi_{\rm gs}({\bf x}_1,{\bf x}_2,{\bf x}_3)d{\bf x}_3;
\label{eq:three_particles_sec_ord_corr_fun}
\end{split}
\end{align}

\noindent 
see Refs.\ \cite{lowd55,alvi12} for details. 

\textcolor{black}{
To proceed further with the Fourier transform, we take the spin orbitals to have a Gaussian-function spatial 
part [see Eq. (\ref{disorb})], that is
\begin{align}
\psi_{j\sigma}(x)=\psi_j(x) \sigma,
\label{disorb2}
\end{align}
\noindent where $\sigma$ denotes the up ($\uparrow$ or $\alpha$) or down ($\downarrow$ or $\beta$) spin. 
As mentioned earlier, the physically relevant quantities are the diagonal parts 
of the two-point correlation function in both real and momentum space \cite{lowd55,alvi12}, i.e.,
}

\begin{align}
\begin{split}
\mathcal{P}^{N=2}_{\rm HL}({\bf x}_1,{\bf x}_2)=
\mathcal{P}_{\rm HL}({\bf x}_1,{\bf x}_1,{\bf x}_2,{\bf x}_2) \label{hlsp}
\end{split}
\end{align}
and
\begin{align}
\begin{split}
\mathcal{G}^{N=2}_{\rm HL}({\bf q}_1^x,{\bf q}_2^x)=
& \frac{1}{4\pi^2}\int_{-\infty}^\infty e^{-i{\bf q}_1^x \cdot ({\bf x}_1 - {\bf x}_1')} 
\int_{-\infty}^\infty e^{-i{\bf q}_2^x \cdot ({\bf x}_2 - {\bf x}_2')}\\
& \times \mathcal{P}_{\rm HL}({\bf x}_1,{\bf x}_1',{\bf x}_2,{\bf x}_2')
d{\bf x}_1 d{\bf x}_1' d{\bf x}_2 d{\bf x}_2', \label{hlmom}
\end{split}
\end{align}

\noindent where here the boldfaced ${\bf q}^x \rightarrow (k,\sigma)$ with $k$ being the momentum along
the $x$ direction; the analytic modeling is performed as a strictly 1D case, unlike the quasi-1D
case of the CI calculations earlier.

\begin{figure}[b]
\includegraphics[width=7.5cm]{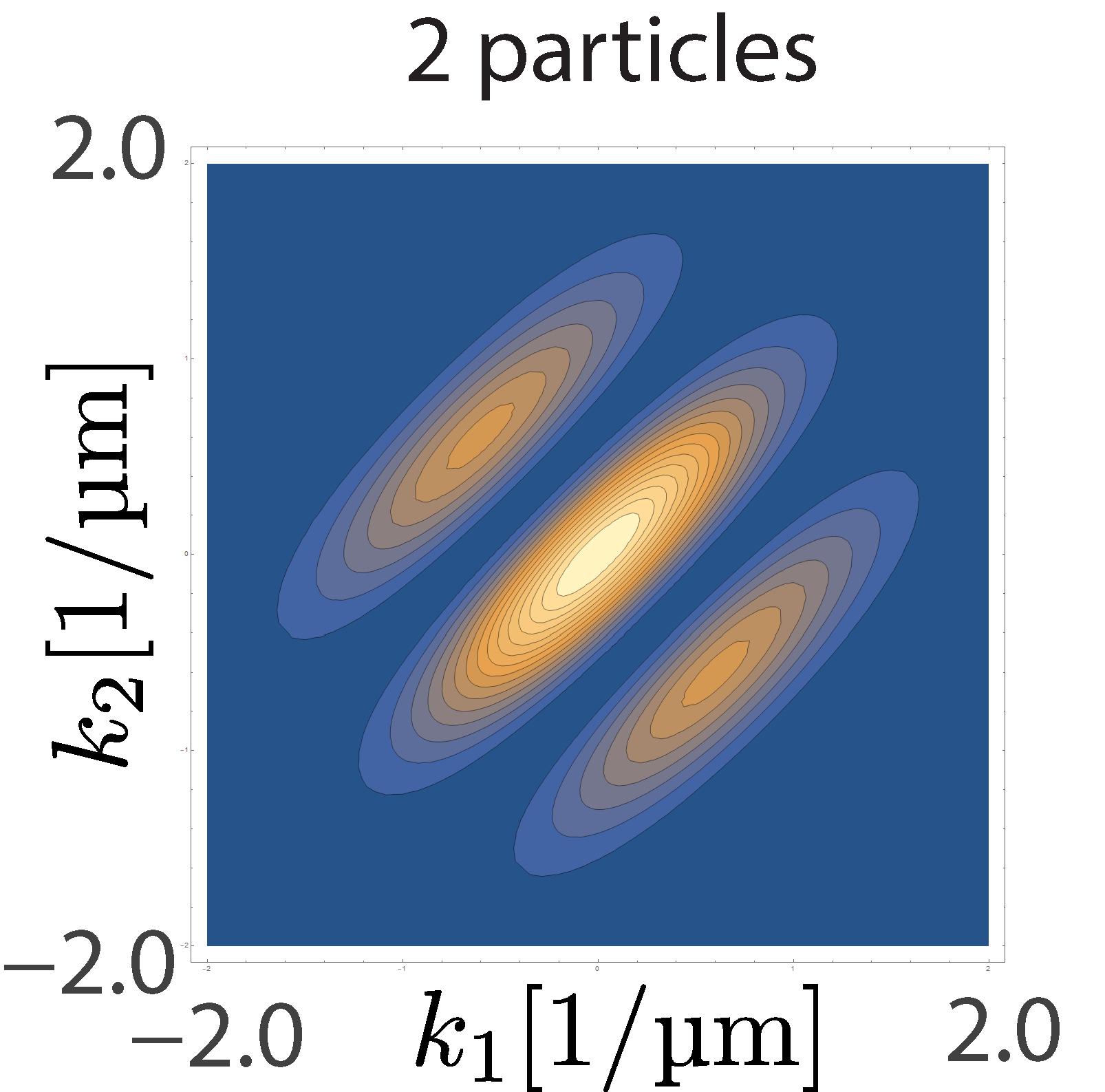} 
\caption{Momentum correlation map for two particles separated by a distance of $2d=4.8$ $\mu$m according to
the analytic model in Sec.\ \ref{anmdtp}. This map was obtained by plotting Eq.\ 
(\ref{eq:2particlesinterference}) with $k_1$ on the horizontal axis and $k_2$ on the vertical axis.
$s=0.71$ $\mu$m.} 
\label{fig1}
\end{figure}

\begin{figure*}[t]
\centering\includegraphics[width=14.cm]{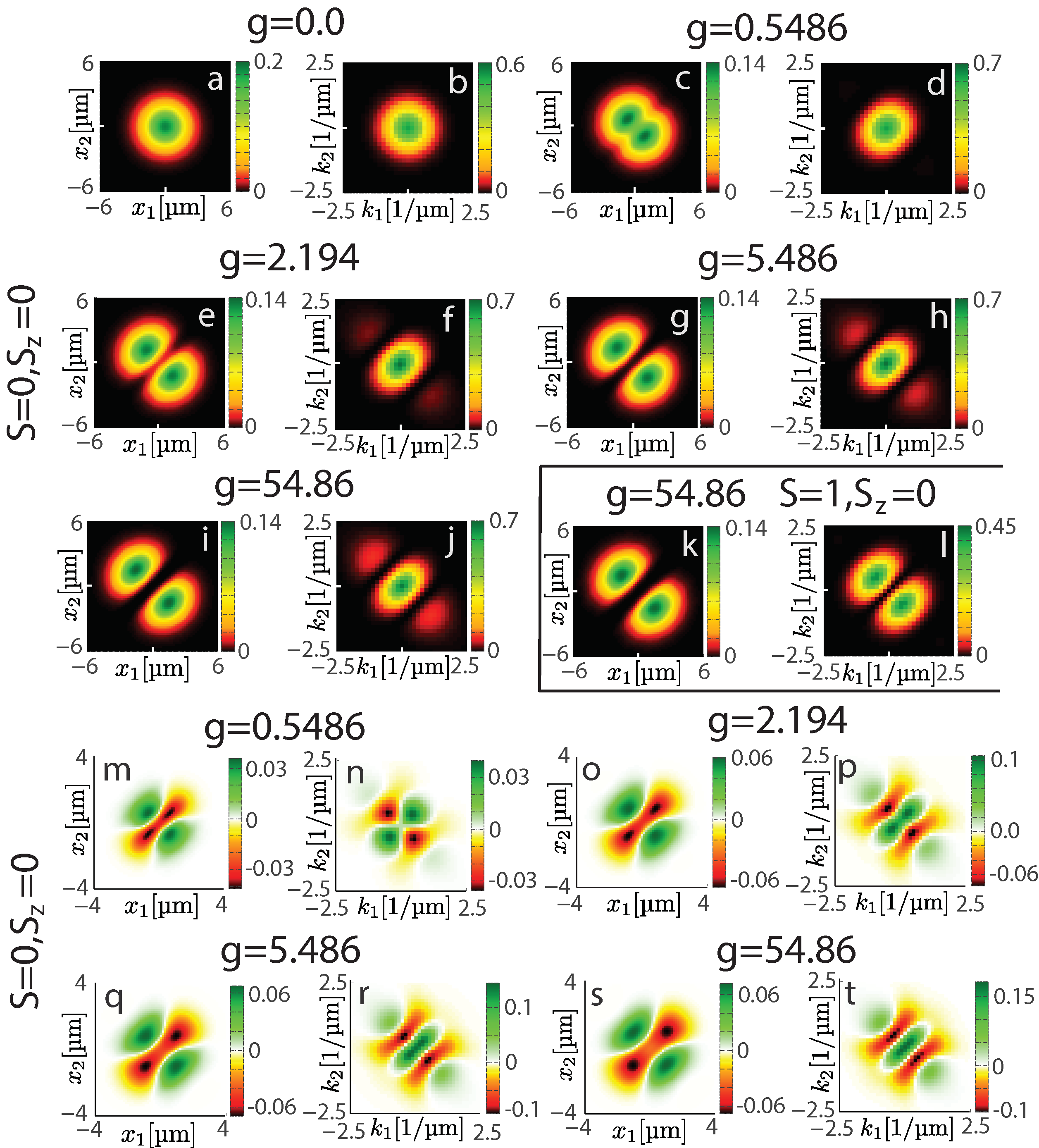}
\caption{
Space (a,c,e,g,i,k) and momentum (b,d,f,h,j,l) two-point correlation maps for CI states of two $^6$Li fermions 
in a single quasi-1D harmonic trap. The interatomic repulsion strength, $g$, in units of $\hbar\omega l_0^2$ 
is indicated in the figure. (m,n), (o,p), (q,r) and (s,t): spatial and momentum noise distributions 
corresponding, respectively, to the (c,d), (e,f), (g,h), and (i,j) correlation maps. Results are shown 
for both the ground-state singlet $S=0$, $S_z=0$ [(a-j) and (m-t)] and the first-excited triplet 
$S=1$, $S_z=0$ (k,l) state. Features amplitudes are given by the color bars on the right of each panel.}
\label{fig1_2}
\end{figure*}

In this step it is pertinent to note that integrals over spin-orbitals with different spins vanish. 
In order to calculate the spin-resolved correlation map, we pick the terms involving the appropriate spin 
orbitals. For instance to calculate the correlation map with down spin for one particle and up spin for 
the second particle, we pick the terms involving $\uparrow^2\downarrow^2$ and $\downarrow^2\uparrow^2$
in Eqs.\ (\ref{hlsp}) and (\ref{hlmom}). For the spin-{\it un}resolved correlation map, we take all spin 
terms into account. For two particles with the Gaussian functions centered at $d_1=-d$ and $d_2=d$, 
we obtain in this way for the spin resolved correlation map the following expression
(we added the superscript $N=2$ to denote the two-particle case illustrated here; see Appendices 
\ref{sr34} and \ref{su34} for $N=3$ and $N=4$):

\begin{align}
{\cal G}_{\rm HL}^{N=2}(k_1\downarrow,k_2\uparrow)=
\frac{4 s ^2 e^{-2 s ^2 \left(k_1^2+k_2^2\right)} \cos ^2[d (k_1-k_2)]}{\pi \mathcal{N}^2_2}
\label{eq:2particlesinterference}
\end{align}  

\noindent which agrees with results found \cite{coul41} in the chemical literature for the case of
the natural H$_2$ hydrogen molecule. An illustration of the diffractive pattern along the cross diagonal
embodied in Eq.\ (\ref{eq:2particlesinterference}) is portrayed in Fig.\ \ref{fig1}.
Here we wish to emphasize that the diffractive 
interference pattern created by the $\cos ^2[d (k_1-k_2)] \propto \left (1+\cos[(d_1-d_2)(k_1-k_2)]\right)$ 
term should be an experimentally detectable signature and it is also the dominant pattern in our CI 
calculations (see Sec.\ \ref{res} below). We also emphasize the presence in Eq.\ 
(\ref{eq:2particlesinterference}) of the cut-off prefactor $e^{-2s^2(k_1^2+k_2^2)}$, which dampens the
constant-amplitude oscillatory behavior of the sinusoidal diffraction term.
The expression for the spin-unresolved correlation map for two particles has the same functional 
form as Eq.\ (\ref{eq:2particlesinterference}). This is a special property of the two opposite spin particles
and for systems with more particles the spin-resolved and spin-unresolved expressions are in general 
different.  

Similar expressions can be derived for the case of three and four fermions in a multi-well potential 
(see Appendices \ref{sr34} and \ref{su34}), with the localized fermions modeled by displaced Gaussian 
functions. A similar diffraction pattern (with an {\it intra\/}-well inter-particle distance 
$2d = d_1-d_2$) develops for two repelling atoms (fermions or bosons) confined in a single well, in the TG 
regime (see below). When analyzing the CI results for ${\cal G}({\bf q}_{1},{\bf q}_{2})$ below, we often 
make the comparison with those from the displaced-Gaussians molecular modeling. 

\begin{figure}[t]
\centering\includegraphics[width=8cm]{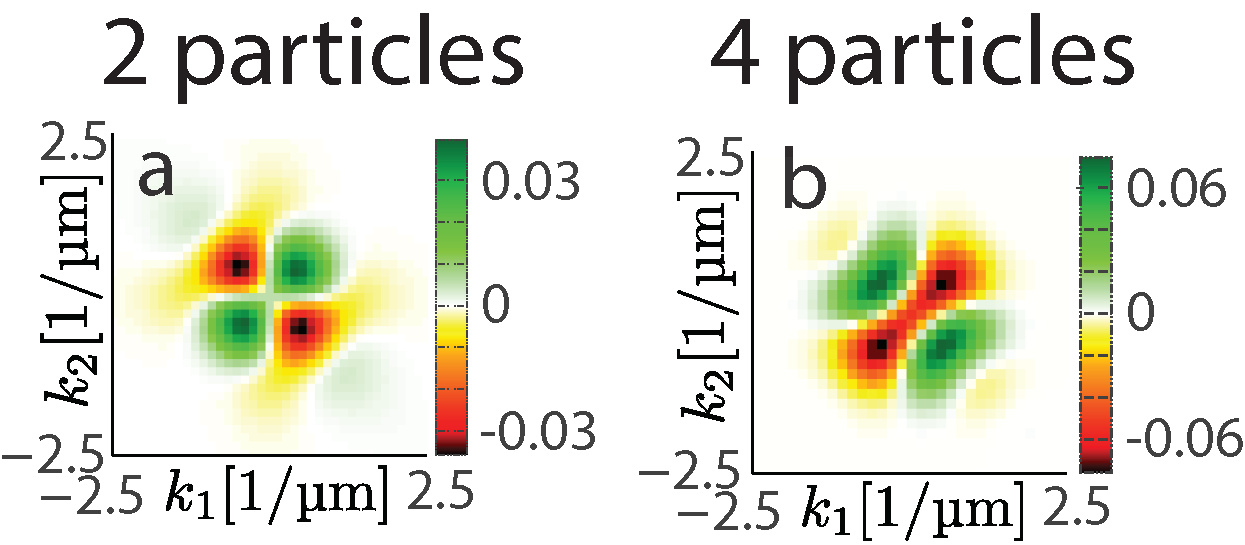}
\caption{
Comparison of CI-calculated two-body momentum correlation map for (a) two and (b) four particles in a quasi-1D 
single well with $g=0.5486$ $\hbar\omega l_0^2$. 
For both system sizes, we find similar characteristic sign-alternation of the momentum correlations in 
adjacent quadrants of the $(k_1, k_2)$ plane, thus supporting the shortsightedness of the two-body momentum 
correlation map. The appearance of the negative correlations (anti-correlations) 
indicates deviations from the Bogoliubov theory.}
\label{fig3_new}
\end{figure}

\begin{figure*}[t]
\centering\includegraphics[width=14cm]{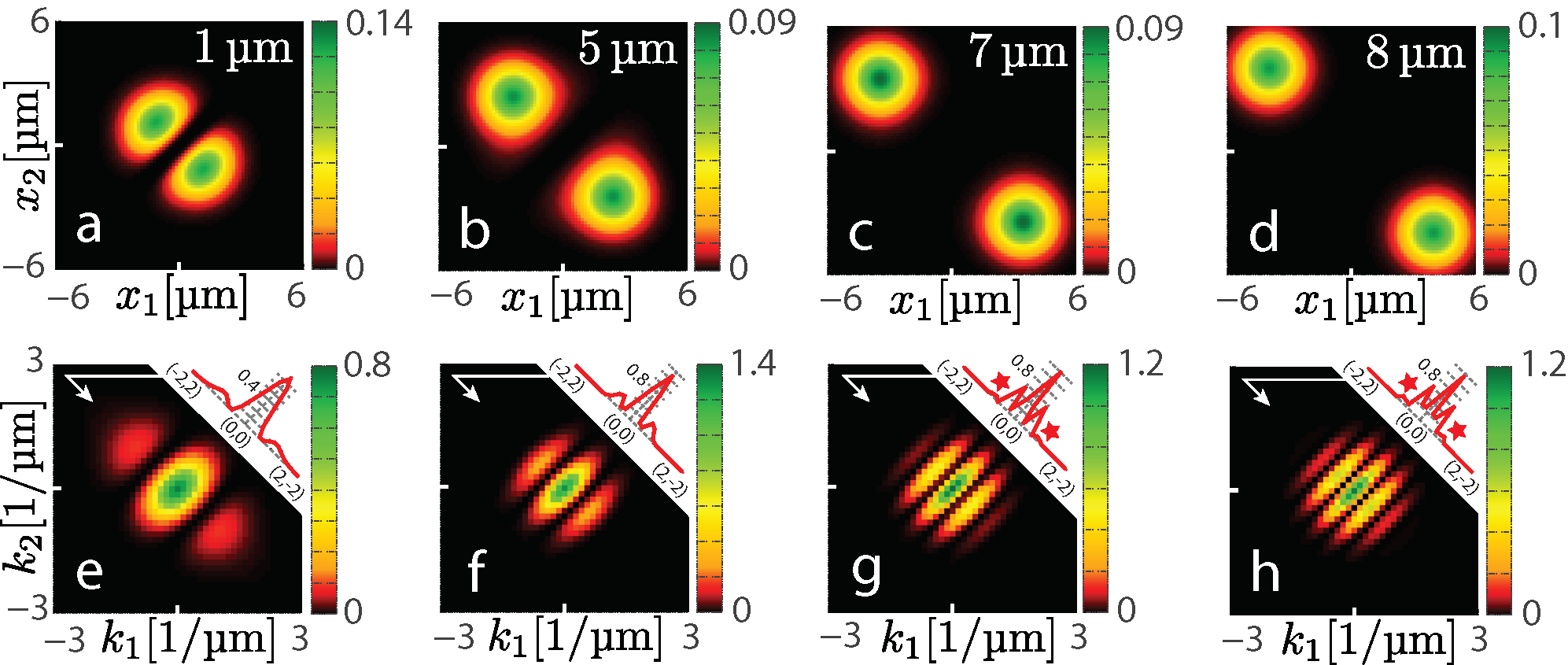}
\caption{
Spin-resolved space (a-d) and momentum (e-h) two-point correlation maps for the CI singlet ground state (with 
$S=0$, $S_z=0$) of two $^6$Li fermions, in quasi-1D double-well traps, with $g=54.86$ $\hbar \omega l_0^2$. 
Four 
different interwell separations $d_w$ are considered. Insets in (e-h) display the variation as a function of 
$k_1$ along the cross-diagonal (the second diffraction peaks are marked by stars in (g,h)).}
\label{fig2_3}
\end{figure*}

\section{Configuration-interaction results}
\label{res} 

\subsection{Two fermions in a single quasi-1D well}

As mentioned earlier, we consider a short-range interparticle repulsion with a Gaussian form
defined in Eq.\ (\ref{tbi}). In Fig.\ \ref{fig1_2} we investigate the evolution with 
increasing repulsion of the two-point momentum correlations [Figs.\ \ref{fig1_2}(b,d,f,h,j,l)]
compared to the corresponding two-point real-space ones [Figs.\ \ref{fig1_2}(a,c,e,g,i,k)] 
in the case of two fermions in a single quasi-1D well. In real space we calculate the CI function 
$\mathcal{P}({\bf w}_1,{\bf w}_2)=\mathcal{P}({\bf w}_1,{\bf w}_1,{\bf w}_2,{\bf w}_2)$ [see Eq.\ 
(\ref{eq:corr_function_def_2})]; in momentum space we calculate ${\cal G} ({\bf q}_1,{\bf q}_2)$ 
[see Eq.\ (\ref{tbmc})]. Because our system is quasi-1D (that is, needle-like shaped along the $x$-direction 
due to the strong confinement in the $y$-direction), it is natural to overlook the variation along the
$y$-direction of the trap and plot the cuts of the previous quantities at $y_1=y_2=0$ and $k_1^y=k_2^y=0$.
This yields the plotted correlation maps for the position ($x_1$, $x_2$) and momentum ($k_1$, $k_2$) 
variables along the long $x$-direction of the trap. The main features in these plots develop along the main 
diagonal (i.e., the line $x_1=x_2$ or $k_1=k_2$, bottom-left to top-right) or the cross-diagonal (i.e., the 
line $x_1=-x_2$ or $k_1=-k_2$).

For the noninteracting $(g=0)$ singlet state, the two-body spatial-correlation density is 
azimuthally uniform having a maximum at $x_1=x_2=0$ [see Fig.\ \ref{fig1_2}(a)]. 
This comes from the fermions with up and down spins occupying the same spatial
$1s$ orbital of the harmonic-oscillator confinement along the $x$ direction. However, as the strength
of the interaction parameter $g$ increases [Figs.\ \ref{fig1_2}(c,e,g,i)], two peaks along the cross-diagonal 
develop and gradually move away from each other. This is reminiscent of the formation of a molecular dimer 
(like the natural H$_2$), often referred to as an ultracold Wigner molecule \cite{yann15}. For large 
$g=5.486$ $\hbar\omega l_0^2$,
a deep valley of almost zero values (black color) develops along the diagonal [Fig.\ \ref{fig1_2}(g)].
For very large $g=54.86$ $\hbar\omega l_0^2$, 
the separation $2d$ between the two peaks saturates and the dimer reaches the 
Tonks-Girardeau regime [Fig.\ \ref{fig1_2}(i)].

This molecule formation is reflected in the evolution of the two-point momentum correlations which
follows the damped diffraction pattern [Eq.\ (\ref{eq:2particlesinterference})] associated with a
Heitler-London wave function. The diffraction pattern develops along the cross-diagonal and the
number of visible diffraction oscillations depends on the distance $2d$ and the spreading of the
product of Gaussian functions ${\cal A}(k_1){\cal A}(k_2)$, with ${\cal A}(k) \propto \exp(-2 k^2 s^2)]$
being the square of the Fourier transform of the space orbital in Eq.\ (\ref{disorb2}) with $d_j=0$.
Characteristically the maximal values of the momentum correlation maps form a ridge along the main diagonal;
such behavior is sometimes termed as ``bunching''. 
For smaller values of $g \leq 0.5486$ $\hbar\omega l_0^2$
[Figs.\ \ref{fig1_2}(b,d)], the separation $2d$ is not large enough to generate secondary maxima along the 
cross diagonal. However, for larger values of $g \geq 2.194$ $\hbar\omega l_0^2$
[Figs.\ \ref{fig1_2}(f,h,j)], the separation $2d$ increases, and a one-oscillation (below, as well as above, 
the main diagonal) diffraction pattern develops which saturates at the Tonks-Girardeau 
limiting regime [the largest value of $g$ considered, Fig.\ \ref{fig1_2}(j)]. For a single well, 
saturation of $2d$ with increasing $g$ limits the number of oscillations in the diffraction pattern due
to the damping factors, and a larger number of diffraction 
oscillations cannot be observed as $g \rightarrow \infty$. 

\begin{figure*}[t]
\centering\includegraphics[width=17.5cm]{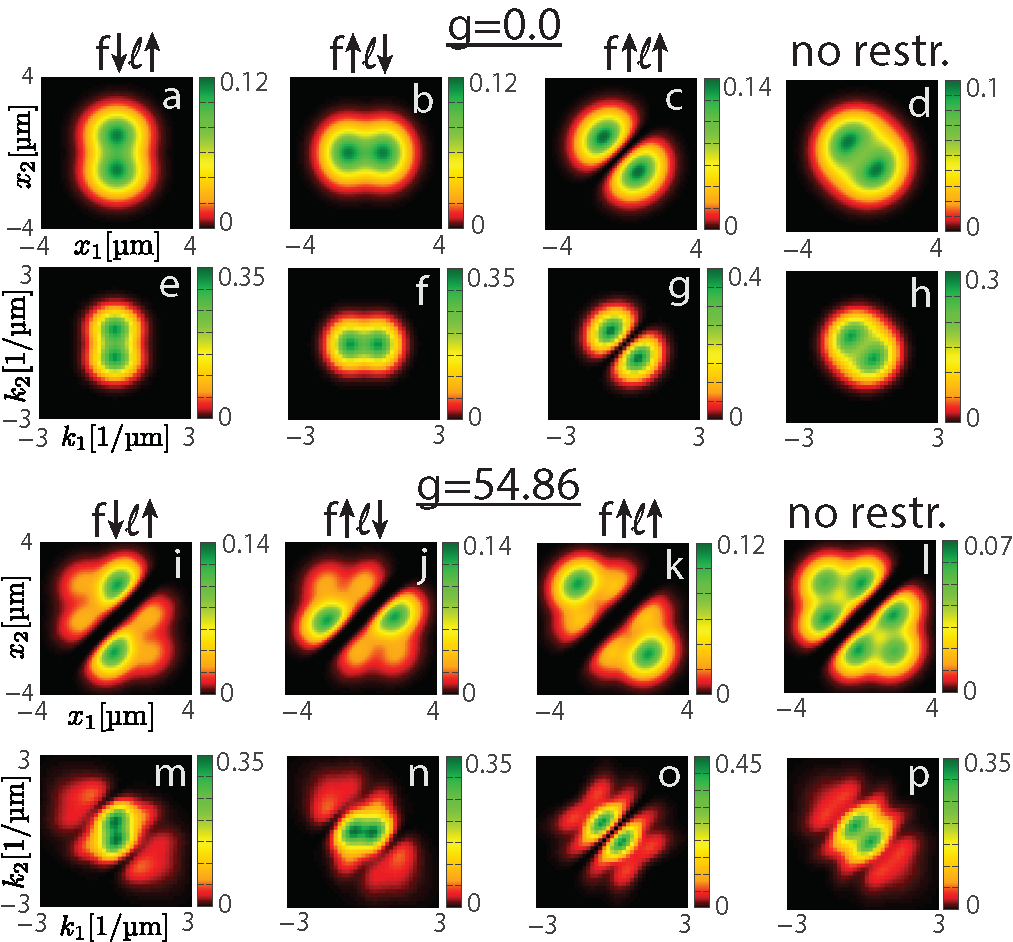}
\caption{
Spin-resolved space (a-d and i-l) and momentum (e-h and m-p) two-point correlation maps for the CI ground state 
(with $S=1/2$, $S_z=1/2$) of three $^6$Li fermions in a quasi-1D single-well trap; non interacting ($g=0$, a-h) 
and strongly repelling ($g=54.86$ $\hbar \omega l_0^2$, i-p). The labels \fdlu , \fuld $\,$ and \fulu 
indicate the spin restriction of the 2D correlation maps. (d,h) and (l,p) give spin-unresolved maps. }
\label{fig3_4}
\end{figure*}

For the triplet state, the short-range repulsion has no influence, and the correlation maps are
independent of $g$. The $g$-independent real-space correlation map [Fig.\ \ref{fig1_2}(k)]
agrees with that of the singlet UCWM 
state near the Tonks-Girardeau limit [large $g$, Fig.\ \ref{fig1_2}(i)], suggesting that the Pauli exclusion 
acts in a similar fashion as a contact repulsion with infinite strength. This is in agreement with the 
well-known mapping between the two-fermion singlet and triplet wave functions referred to as 
``fermionization'', observed also experimentally \cite{joch12}. The 
corresponding momentum correlation map 
for the triplet [Fig.\ \ref{fig1_2}(l)], however, is drastically different compared to that of the singlet 
state [Fig.\ \ref{fig1_2}(j)]. In particular, the momentum correlation map
exhibits a deep trough (colored black) along the main diagonal instead of a ridge (colored green);
such trough formation is sometimes termed ``anti-bunching''. This
trough denotes a vanishing of the probability for finding two fermions with parallel spins having the
same momentum, a property imposed by the Pauli principle in momentum space.  

\textcolor{black}{
It is rewarding to note that the analytic modeling yields results in full agreement with the CI result in
Fig.\ \ref{fig1_2}(l). Indeed the two-body momentum correlations for the Heitler-London triplet, built out of 
two displaced Gaussian space orbitals (positioned at $d_1=-d$ and $d_2=d$), are given by
\begin{align}
\mathcal{G}_{\rm HL,t}^{N=2}(k_1 \uparrow,k_2 \uparrow) \propto
\frac{4 s ^2 e^{-2 s ^2 \left(k_1^2+k_2^2\right)} \sin ^2(d (k_1-k_2))}{\pi }.
\label{anhlt}
\end{align} 
\noindent It is apparent that the term $\sin ^2(d (k_1-k_2))$ in Eq.\ (\ref{anhlt}) reproduces the deep trough
visible in the CI correlation map [see Fig.\ \ref{fig1_2}(l)] along the main diagonal $(k_1=k_2)$.} 

\begin{figure*}[t]
\centering\includegraphics[width=14cm]{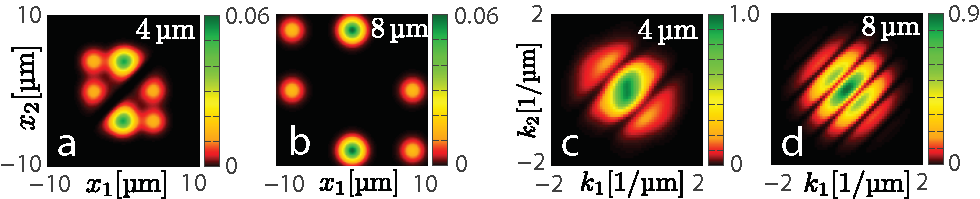}
\caption{
Spin-resolved space (a,b) and momentum (c,d) two-point correlation maps for the CI ground state (with 
$S=1/2$, $S_z=1/2$) of three $^6$Li fermions in quasi-1D triple-well traps ($g=54.86$ $\hbar \omega l_0^2$). 
Two different interwell separations $d_w$ are considered.}
\label{fig4_5}
\end{figure*}

From the two-point correlation maps, one may extract the often used \cite{bouc16,altm09,altm04} 
corresponding noise distributions [Fig.\ \ref{fig1_2}(m-t)]; the two-point noise distributions are obtained by 
subtracting the product of the corresponding one-point momentum correlations; see 
Eqs.\ (\ref{pn}) and (\ref{gn}). These noise distributions show both positive and negative values, with 
the negative ones corresponding to the vanishing probability troughs in the correlation maps proper.
In the case of real-space plots [Figs.\ \ref{fig1_2}(m,o,q,s)], the noise distributions again 
reveal the progressively increasing separation of two positive peaks (colored green) along the cross 
diagonal, which corresponds to the formation of a UCWM. For the momentum plots [Figs.\ \ref{fig1_2}(n,p,r,t)],
it is remarkable that for the weak repulsion value $g = 0.5486$ $\hbar \omega l_0^2$ [Fig.\ \ref{fig1_2}(n)],
our noise distributions closely resembles the QBEC square-shaped pattern $(+,-,+,-)$ measured for a system 
comprised of a large number of 1D bosons \cite{bouc16}. For even stronger $g$'s [Figs.\ \ref{fig1_2}(p,r,t)], 
close to the TG regime, our noise maps display a more complex shape that reflects the oscillations in the 
corresponding diffraction pattern of the two-point momentum correlations, that is two negative 
areas (red color) enclosed by three positive areas (green color). 

\textcolor{black}{ 
Before presenting our results for multi-well systems, we illustrate in Fig.\ \ref{fig3_new} the 
shortsightedness of the two-body momentum correlations by comparing 
the two-body momentum noise maps for two [Fig.\ \ref{fig3_new}(a)] and four 
[Fig.\ \ref{fig3_new}(b)] particles confined in a single well, for a repulsion strength 
$g = 0. 5486 \hbar \omega l_0^2$ [see momentum noise map in Fig.\ \ref{fig1_2}(n)]; for additional information 
about the shortsightedness of the momentum correlation function, see text 
following Eq.\ (\ref{eq:spinresolved4particles}) in Appendix \ref{sr34}. Comparison of the noise maps
in Fig.\ \ref{fig3_new}, reveals that these show similar characteristic sign-alternations portraying 
opposite-momentum correlations and anticorrelations, as predicted \cite{bouc12} and, more recently, observed 
experimentally \cite{bouc16}. As we noted in the Introduction, the appearance of such characteristics in the 
two-body momentum noise correlations is a signature of deviations from the time-honored Bogoliubov theory 
\cite{bouc12,bouc16}, whose treatment necessitates many-body theories beyond the mean-field approximation. 
Underlying the persistent appearance of these characteristics in few particle quasi-1D systems of variable size 
(see Fig.\ \ref{fig3_new}) is the aforementioned shortsightedness of the two-body momentum correlations.}

\textcolor{black}{
These findings support our suggestion that investigations of few-body systems could be used to shed light on 
experimental observations pertaining to certain complex many-body properties (such as the effect of 
interparticle interactions of variable strength on the nature of quantum liquids, 
including deviations from the Bogoliubov 
theory in quasi-1D systems in the QBEC regime and for stronger repulsive interactions, that is the TG regime) 
even when such experiments are carried on larger systems (see, e.g., Ref. \cite{bouc16}). 
}

\subsection{Two fermions in a quasi-1D double well}

To gain further insight into the trends generated through varying the separation between the two 
high-probability peaks in the real-space two-point correlation maps, we display in Fig.\ \ref{fig2_3} spatial 
and momentum correlation maps for the CI singlet ground state of two $^6$Li atoms confined in quasi-1D 
double-well traps at different interwell separations $d_w=2d=1$, 5, 7, and 8 $\mu$m. An important observation is
that the pair of maps for the smallest separation $d_w=1$ $\mu$m [see Figs.\ \ref{fig2_3}(a,e)] closely 
resembles those of the two fermions in a single well near the Tonks-Girardeau regime [see Figs.\ 
\ref{fig1_2}(i,j)]. This further supports the interpretation of the Tonks-Girardeau regime as a special limit 
of the more general Wigner-molecule approach, which extends also to 2D and 3D systems \cite{yann04}. As the 
interwell separation increases, from 5 [Figs.\ \ref{fig2_3}(b,f)] to 7 $\mu$m [Figs.\ \ref{fig2_3}(c,g)] and
8 $\mu$m [Figs.\ \ref{fig2_3}(d,h)], an additional diffraction oscillation gradually emerges, becoming clearly 
visible for the separation of 8 $\mu$m. 

\subsection{Three fermions in a quasi-1D single and triple wells}

Fig.\ \ref{fig3_4} displays the evolution of the spatial and momentum two-point correlations for the $S=1/2$, 
$S_z=1/2$ ground-state of $N=3$ $^6$Li atoms in a single-well trap for the non-interacting $(g=0)$  
[Figs.\ \ref{fig3_4}(a-h)] and the strongly repelling ($g=54.86$ $\hbar \omega l_0^2$) case 
[Figs.\ \ref{fig3_4}(i-p)]. Furthermore, both cases of spin-resolved 
[Figs.\ \ref{fig3_4}(a,b,c,e,f,g) and Figs.\ \ref{fig3_4}(i,j,k,m,n,o)] and with no-spin restriction 
[Figs.\ \ref{fig3_4}(d,h) and Figs.\ \ref{fig3_4}(l,p)] are presented. 
In interpreting these maps, we can use the spin-resolved conditional probability distribution function defined 
in Refs.\ \cite{li07,li09} and in Eqs.\ (6) and (7) of Ref.\ \cite{yann15}. 
First, we invoke the spin-resolved two-point anisotropic correlation function.
The spin-resolved two-point anisotropic correlation function is defined as
\begin{eqnarray}
&& P_{\sigma\sigma_f}({\bf r}, {\bf r}_f)= \nonumber \\
&& \langle \Phi_{\rm CI}^N |
\sum_{i \neq j} \delta({\bf r} - {\bf r}_i) \delta({\bf r}_f - {\bf r}_j)
\delta_{\sigma \sigma_i} \delta_{\sigma_f \sigma_j}
|\Phi_{\rm CI}^N\rangle.
\label{tpcorr}
\end{eqnarray}
Using a normalization constant 
${\cal N}(\sigma,\sigma_f,{\bf r}_f) = 
\int P_{\sigma\sigma_f}({\bf r}, {\bf r}_f) d{\bf r}$,
we further define a related spin-resolved conditional probability distribution (CPD) as
\begin{equation}
{\cal P}_{\sigma\sigma_f}({\bf r}, {\bf r}_f) =
P_{\sigma\sigma_f}({\bf r}, {\bf r}_f)/{\cal N}(\sigma,\sigma_f,{\bf r}_f).
\label{cpd}
\end{equation}

The label ``$f \downarrow$'' in ``$f \downarrow \ell \uparrow$'', corresponds to a selected observation 
(``fixed'', or ``f'') point, with  the arrow denoting the chosen spin direction at that observation point. For 
that selected observation (``fixed'') point on the $x_1$ (or $k_1$) axis, corresponding to particle ``1'',  
we search (``look for'', or ``$\ell$'') at all points along the $x_2$ (or $k_2$) axis, corresponding to particle
''2'',  with a spin direction $\uparrow$, and record in the map the probabilities of finding particle ``2'' 
with the specified spin direction at these points. Repeating this process for all values along the $x_1$ axis 
(that is, all observation points) completes the interpretation of the label ``\fdlu'' in the correlation maps. 
To reiterate -- the physical meaning of the notation ``\fdlu'', ``\fuld'', ``\fulu''
is based on the fact that a conditional probability can be
extracted from the correlation maps by fixing the indices of one particle, i.e., spin and position. Indeed the 
cuts in the correlation maps defined by $x_1={\rm constant}$ ($k_1={\rm constant}$) portray the conditional 
probability of finding a second particle with predetermined spin at $x_2$ ($k_2$) assuming that the first 
particle with given spin is fixed at $x_1={\rm constant}$ ($k_1={\rm constant}$).  

To facilitate understanding of the spin-resolved maps in Fig.\ \ref{fig3_4}, we mention that for $g=0$ the 
many-body configuration is $1s^2 1p$, i.e., there are two spin-up fermions occupying the $1s$ and $1p$ 
orbitals and one spin-down fermion occupying again the $1s$ orbital. For the strong $g=54.86$ 
$\hbar \omega l_0^2$, the appropriate spin function for a linear Wigner molecule of three localized fermions 
is \cite{li07}
$(2|\uparrow\downarrow\uparrow\rangle - 
|\uparrow\uparrow\downarrow\rangle -
|\downarrow\uparrow\uparrow\rangle)/\sqrt{6}$. 
For the noninteracting case, our CI calculations give double-peaked space and momentum  correlation maps  
(\fdlu and \fuld) that reflect the presence of the $1p$ orbital
[Figs.\ \ref{fig3_4}(a,e,b,f)]. Fixing a spin-up and looking for the other spin-up (\fulu) exhibits a
valley of vanishing probabilities along the main diagonal; this is a reflection of the Pauli fermion
statistics in both the space and momentum correlations [Figs.\ \ref{fig3_4}(c,g)]. The spin-unresolved 
correlations [Figs.\ \ref{fig3_4}(d,h)] can be understood as the sum of the three spin-resolved
ones.

The UCWM case when $g=54.86$ $\hbar \omega l_0^2$ exhibits structures in real-space maps [Figs.\ 
\ref{fig3_4}(i-l)] associated with the three localized fermions, i.e., 
a total of six peaks. For the spin-resolved maps [Figs.\ \ref{fig3_4}(i-k)], a pair of peaks is 
stronger, as follows from the UCWM spin function listed
above (see the coefficient 2). Unlike the noninteracting case, the valleys of vanishing probabilities along 
the main diagonal are present for all three spin-resolved maps [Figs.\ \ref{fig3_4}(i-k)]; 
this is due to the fact that the three fermions do not 
overlap because they are well-localized by the strong repulsion. The momentum maps [Figs.\ \ref{fig3_4}(m-p)], 
however, are not as revealing as the space maps concerning the particle localization. 
Indeed, qualitatively, the main pattern in these maps is similar to that found for two fermions 
[see Fig.\ \ref{fig1_2}]. Namely, there is a damped diffraction pattern along the cross diagonal exhibiting an 
oscillation (below, as well as above, the main diagonal) with one minimum and
one secondary maximum; see pairs of narrow black troughs in Figs.\ \ref{fig3_4}(m,n,p). This pair of troughs 
is less prominent for the \fulu case [Fig.\ \ref{fig3_4}(o)] where a strong valley of vanishing probability 
develops along the main diagonal due to the Pauli exclusion principle. Naturally, there are still significant
quantitative differences between the maps in Figs.\ \ref{fig3_4}(m-p) and the maps in Fig.\ 
\ref{fig1_2}, which could be explored experimentally.

To explore further the diffraction pattern for three localized fermions, we display in Fig.\ \ref{fig4_5}
spin-resolved (\fdlu) real-space and momentum maps for three fermions in a triple-well trap for two
different interwell separations $d_w=2d=4$ $\mu$m and $d_w=2d=8$ $\mu$m in the UCWM case ($g=54.86$ 
$\hbar \omega l_0^2$). As noted earlier
for two fermions, increasing the separation enhances the prominent features described in Fig.\ \ref{fig3_4}
for the three fermions in a single well. In particular, the patterns for the real-space correlations 
are enhanced versions of the pattern in Fig.\ \ref{fig3_4}(i). The momentum correlation map for $2d=4$ $\mu$m 
[Fig.\ \ref{fig4_5}(c)] shows a single diffraction oscillation along the cross diagonal. However, the momentum 
correlation map for $2d=8$ $\mu$m [Fig.\ \ref{fig4_5}(d)] shows a well-developed second diffraction 
oscillation, in agreement with the analytic formula of the simple model listed in the Appendix \ref{sr34}.   
 
\section{Comparison of analytical predictions with CI results}
\label{comp}

\begin{figure*}[t]
\centering
\includegraphics[width=14cm]{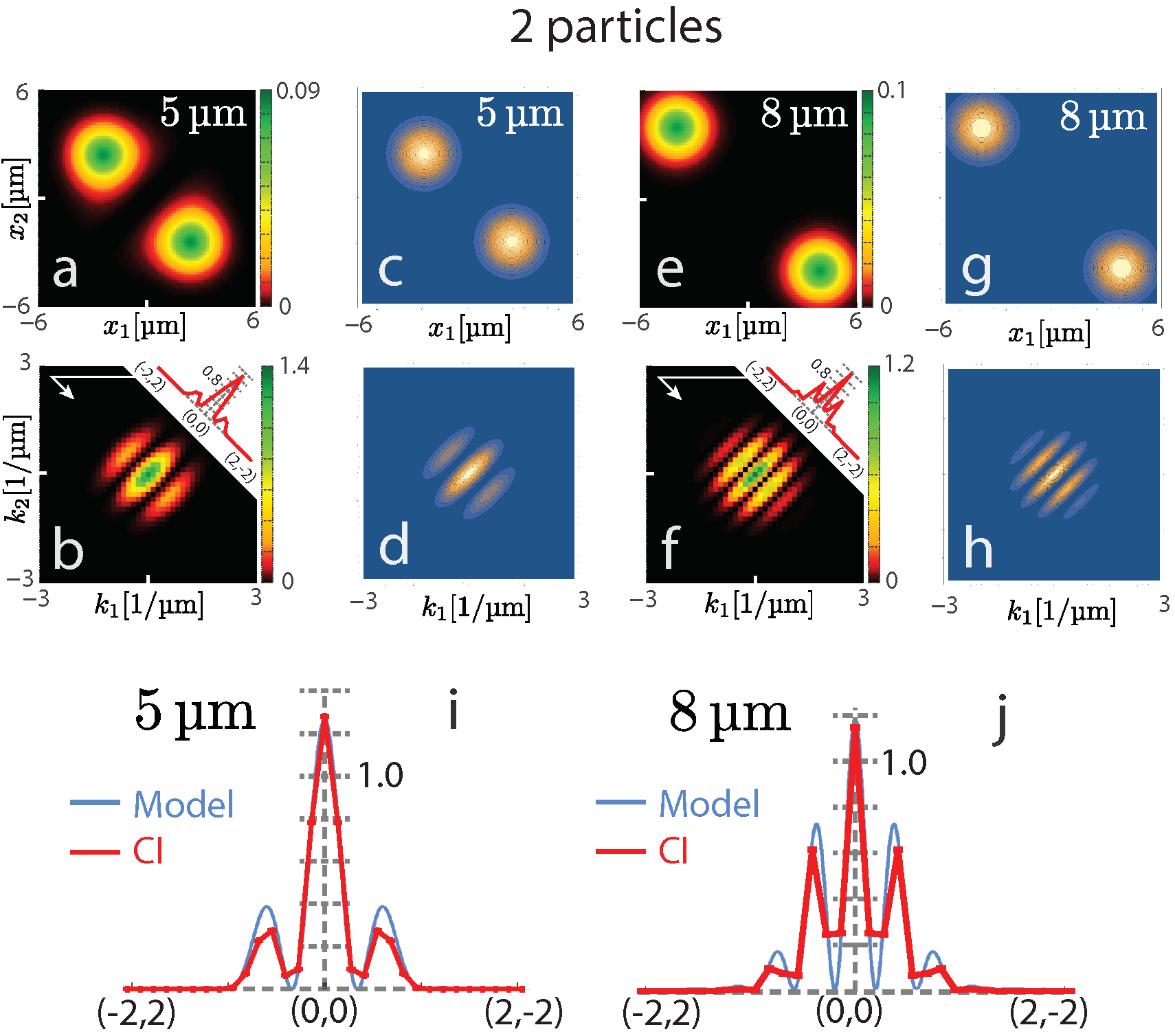}
\caption{Comparing the spin-resolved correlation map predictions from the analytic model with the CI results 
calculated for the singlet ground state (with $S = 0$, $S_z = 0$) of two $^6$Li fermions, interacting  with 
$g = 54.86$ $\hbar \omega l_0^2$, in a quasi-1D double-well trap (see Fig.\ \ref{fig2_3}).
The inter-well distance $d_w$ in $\mu$m is indicated in the figure panels. 
The first row gives the two-point spatial correlation maps, and the second row shows the two-point momentum 
correlation maps. The first (a,b) and third (e,f) columns correspond to the CI results, and the second (c,d) 
and  fourth (g,h) display the analytic results; the insets in (b) and (f) show cuts in the momentum correlation 
maps along the cross-diagonal. In (i) and (j) we show, respectively, a cut through the cross-diagonal in the CI 
and analytic momentum-correlation maps, calculated for $d_w = 5$ $\mu$m (b,d), and for $d_w = 8$ $\mu$m (f,h); 
the analytic results in (i) and (j) (blue curves) were matched at their maximum value to the maximum in the 
corresponding CI (red curves) calculated momentum correlation. In the analytic formulas, $s=0.91$ $\mu$m for
both distances $d_w=5$ $\mu$m and $d_w = 8$ $\mu$m.}
\label{fig:two_particle_MMA_CI_comparison}
\end{figure*}

\begin{figure*}[b]
\centering
\includegraphics[width=14cm]{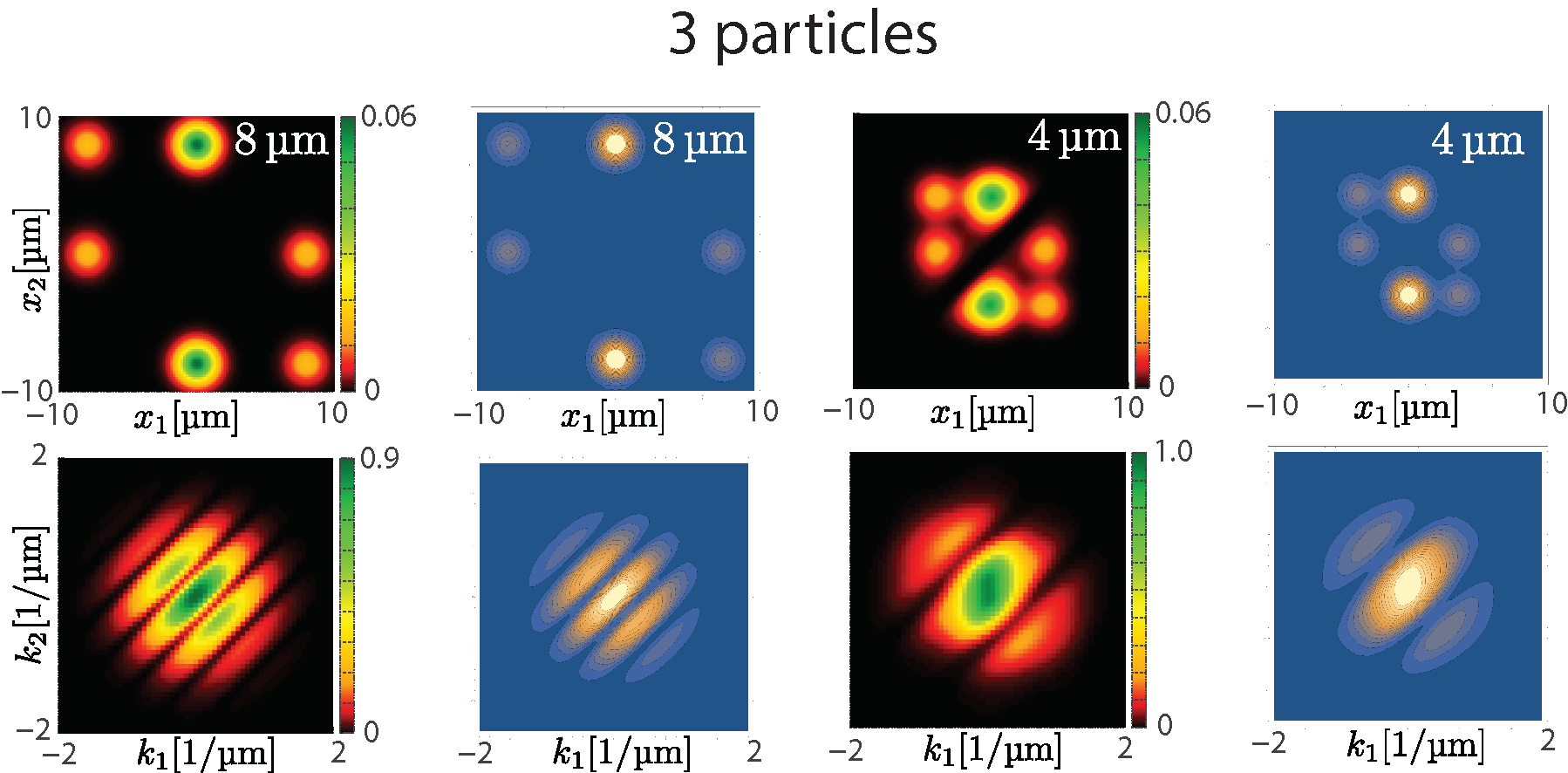}
\caption{Comparing the spin-resolved real-space (top row) and momentum-space (bottom row) correlation-map
predictions from the analytic model with the CI results calculated for the ground state
(with $S = 1/2$, $S_z = 1/2$) of three $^6$Li fermions in a quasi-1D triple-well trap with interaction strength 
of $g=54.86$ $\hbar \omega l_0^2$ (see Fig.\ \ref{fig4_5}). 
The first and third column represent the CI results, the second and fourth the 
analytic results. The interwell distances $d_w$ (evenly spaced wells) are specified in the panels.
In the analytic formulas, $s=0.91$ $\mu$m for both interwell distances.}
\label{fig:three_particle_MMA_CI_comparison}
\end{figure*}

\begin{figure*}[t]
\centering
\includegraphics[width=14cm]{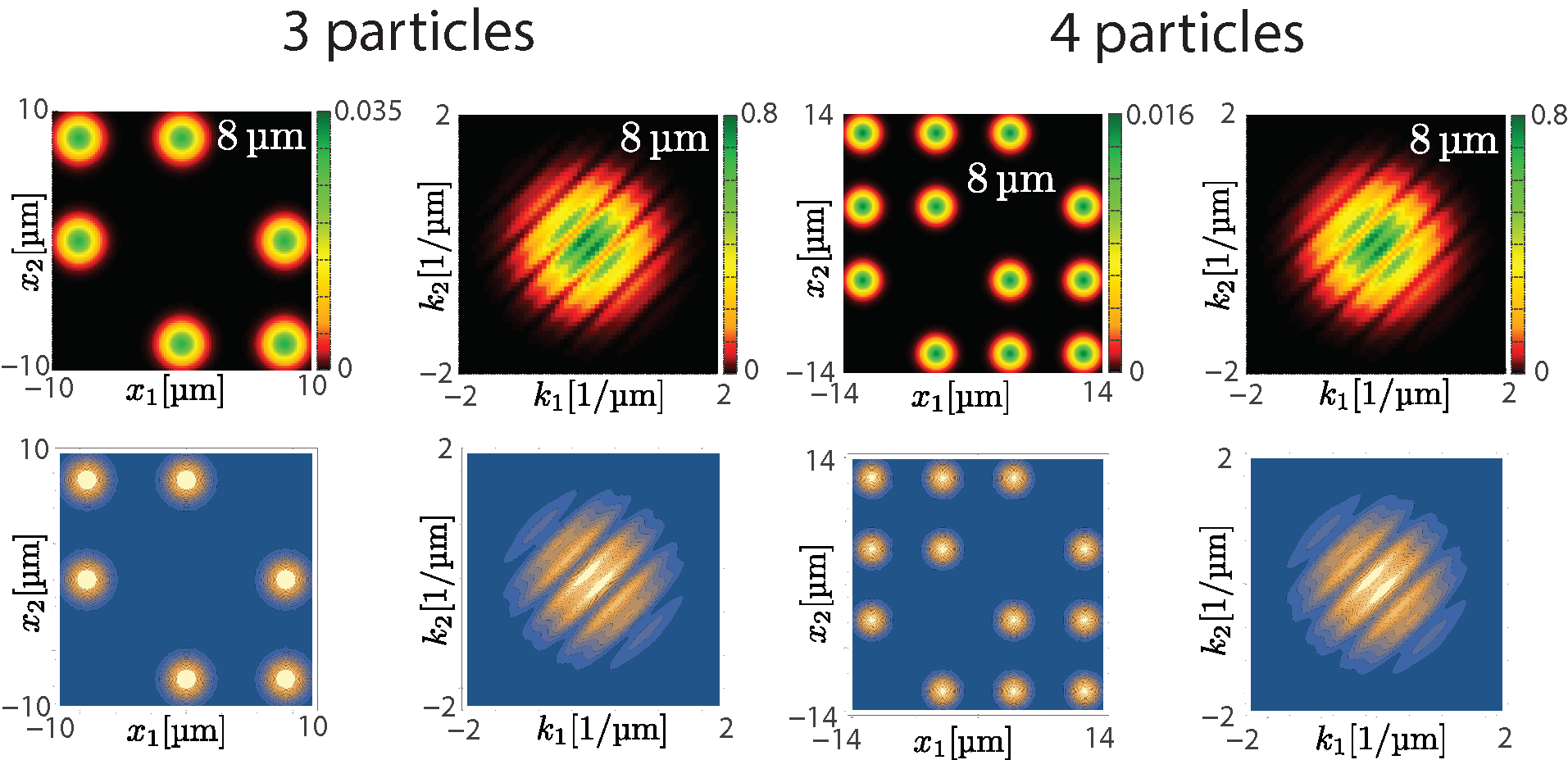}
\caption{Comparing the spin {\it un\/}resolved two-body correlation map predictions from the analytic model 
with the spin-unresolved CI results for the ground state of three $^6$Li fermions in a quasi-1D triple-well 
trap, and four $^6$Li fermions in a quasi-1D quadruple-well trap, with interaction strength of $g=54.86$
$\hbar \omega l_0^2$. The inter-well 
distance $d_w$ (evenly spaced wells) in $\mu$m is indicated in the figures. The first and second column 
represent the three particle results, the third and fourth column represent the four particle results. The top 
row shows the CI results, the bottom row shows the analytic predictions. For each distance we show both the 
real space-correlation function (left) and the momentum-space one (right). In the analytic formulas,
$s=0.91$ $\mu$m for both the cases of three and four $^6$Li fermions.
}
\label{fig:three_four_particle_MMA_CI_comparison}
\end{figure*}

The success of the analytic modeling is evidenced by comparing analytical predictions with the 
{\it ab-initio\/} CI  numerical results. In  Fig.\ \ref{fig:two_particle_MMA_CI_comparison}, 
Fig.\ \ref{fig:three_particle_MMA_CI_comparison}, and Fig.\ \ref{fig:three_four_particle_MMA_CI_comparison}, 
we compare the CI-calculated correlation maps with the correlation maps obtained from the analytical 
expressions (see Sec.\ \ref{anmdtp} and Appendices \ref{nm2p}, \ref{sr34}, and \ref{su34}) for two, three, 
and four particles in double, triple, and quadruple well confinements, respectively. The agreement between 
both methods is excellent. We note here that the model used here (localized Gaussian functions with small 
overlap, and the Heisenberg Hamiltonian) becomes more complicated for smaller interwell distances. 
Interestingly, for the cases that we have investigated here the analytical expressions that we have derived 
from our model predict adequately, at least qualitatively, the features found through the microscopic CI 
calculations.  

\section{Conclusions} 
\label{conc}

In this paper we have explored systematically the characteristics of spin-resolved spatial and momentum-space
correlations and noise distributions for two, three, and four ultracold fermionic atoms trapped in single and
multiple wells; see also Appendices \ref{nm2p}, \ref{sr34}, and \ref{su34}. These investigations aim at 
gaining insights into the quantum states of different phases of ultracold matter and the nature of trapped 
multiple-ultracold-atoms molecule-like assemblies, and providing fingerprinting guidance for experiments, 
particularly ones with a few optically trapped, deterministically prepared and spin resolved, ultracold 
fermionic atoms. 

Using full configuration-interaction exact-Hamiltonian diagonalization, we have evaluated and investigated 
two-point spatial and momentum-space correlations and noise distributions for the entire range of interatomic
contact repulsions and interwell distances, exploring the transition from a 
\textcolor{black}{
noninteracting assembly to the quasi Bose-Einstein condensate and then to the}
Tonks-Girardeau regime. A main result emerging from our numerical simulations using the exact many-body CI 
wavefuncions is a damped oscillatory diffraction behavior of the two-point momentum correlations and noise 
distributions, agreeing with our analytical model results for multiple ultracold fermionic atoms trapped in 
single and multiple wells. 

\textcolor{black}{
Furthermore, the two-body momentum correlation and noise distributions are found to exhibit shortsightedness, 
with the main contribution coming from nearest-neighboring particles. This suggests that investigations of 
two-body (and possibly higher-order) momentum correlations in few-particle confined systems could be employed 
in the interpretation of studies carried on larger particle systems. We illustrated this approach for quasi-1D 
few-fermion systems with intermediate repulsive interactions which yielded two-body momentum noise correlations 
exhibiting opposite-momentum correlations and anticorrelations at small momenta, which closely resemble those 
predicted \cite{bouc12} and measured \cite{bouc16} for a system comprised of a large number of 1D
bosons in the QBEC regime. These studies address deviations from the celebrated  Bogoliubov theory of quantum 
liquids. Moreover, a more complex characteristic pattern is predicted by our calculations in the 
Tonks-Girardeau regime.}
The treatment developed here, which incorporates the effects of interatomic interactions in the
two-body momentum and spatial correlations, goes beyond the Hanbury Brown and Twiss interferometry, where the
free-particle statistics brings about bunching (fermions) versus anti-bunching (bosons) behavior [4,5]; in 
this context compare Figs.\ 2(j) and 2(l) for the two interacting-particles singlet and triplet states, 
respectively.

\begin{acknowledgments} 
This work has been supported by a grant from the Air Force Office of Scientific Research (USA) under 
Award No. FA9550-15-1-0519. Calculations were carried out at the GATECH Center for Computational Materials 
Science.
\end{acknowledgments} 

\appendix

\section{Analytic modeling: Noise maps for two particles}
\label{nm2p}

To illustrate the formation of the patterns seen in the noise maps, we outline in this appendix the calculation 
of the noise distribution for two particles separated by a distance $2d$. 
The calculation of the two-body correlation function proceeds as described in Sec.\ \ref{anmdtp}, with the 
one-body correlation function obtained by applying Eq.\ (\ref{eq:corr_function_def}), or by evaluating directly 
from the many-body wavefunction as described in Ref.\ \cite{lowd55}. We have derived the analytic expressions
using the algebraic computer program MATHEMATICA. In general, for the noise maps for $N=2$ and for the
two-body correlation functions for $N>2$, these expressions are too long and complicated to be reproduced in 
print. For simplicity, in this Appendix and in Appendices \ref{sr34} and \ref{su34}, we present the 
analytic results for the case of strongly localized particles when the overlaps $S_{ij}$ between adjacent
space orbitals can be neglected; in this case, ${\cal N}_2 \approx 1$.

Having obtained the one- and two-body correlation functions, the noise maps can be obtained by applying 
Eqs.\ (\ref{pn}) and (\ref{gn}). Setting $d_1=-d$ and $d_2=d$, the needed product expressions for the one-body 
correlation function in real and momentum space are

\begin{widetext}
\begin{align}
\rho(x_1,x_1)\rho(x_2,x_2)=
{\cal C}(s)\frac{e^{-\frac{(d+x_1)^2}{2 s ^2}} \left(e^{\frac{2 d x_1}{s ^2}}+
1\right)}{2 \sqrt{2 \pi } s }\frac{e^{-\frac{(d+x_2)^2}{2 s ^2}} 
\left(e^{\frac{2 d x_2}{s ^2}}+1\right)}{2 \sqrt{2 \pi } s }
\end{align}
\noindent and
\begin{align}
\tau({k}_1,{k}_1)\tau({k}_2,{k}_2)=
{\cal C}(s){\frac{2}{\pi }} s^2  e^{-2 (k_1^2 + k_2^2) s ^2}
\end{align}

\noindent and the expressions for the two-body correlation functions are:

\begin{equation}
\begin{split}
\mathcal{P}^{N=2}({x_1},{x_2})=
{\cal C}(s) \frac{e^{-\frac{2 d^2+2 d (x_1+x_2)+x_1^2+x_2^2}{2 s ^2}} \left(e^{\frac{d x_1}{s ^2}}
+e^{\frac{d x_2}{s ^2}}\right)^2}{4 \pi  s ^2}
\end{split}
\end{equation}

\begin{align}
\mathcal{G}^{N=2}({k_1},{k_2})=
{\cal C}(s)\frac{4 s ^2 e^{-2 s ^2 \left(k_1^2+k_2^2\right)} \cos ^2 \big( d (k_1-k_2) \big)}{\pi }
\end{align}

\noindent which after subtraction of the one-body terms yield the desired expressions for the noise maps 
[see Eqs.\ (\ref{pn}) and (\ref{gn})]

\begin{equation}
\begin{split}
{\mathcal{P}}^{N=2}_\mathcal{N}(x_1,x_2)=
\mathcal{P}^{N=2}(x_1,x_2)-\rho(x_1,x_1)\rho(x_2,x_2)
\label{anpn}
\end{split}
\end{equation}

\begin{align}
{\mathcal{G}}^{N=2}_\mathcal{N}({k}_1,{k}_2)=
\mathcal{G}^{N=2}(x_1,x_2)-\tau(x_1,x_1)\tau(x_2,x_2)
\label{angn}
\end{align}
\end{widetext}

\begin{figure*}[t]
\centering\includegraphics[width=14cm]{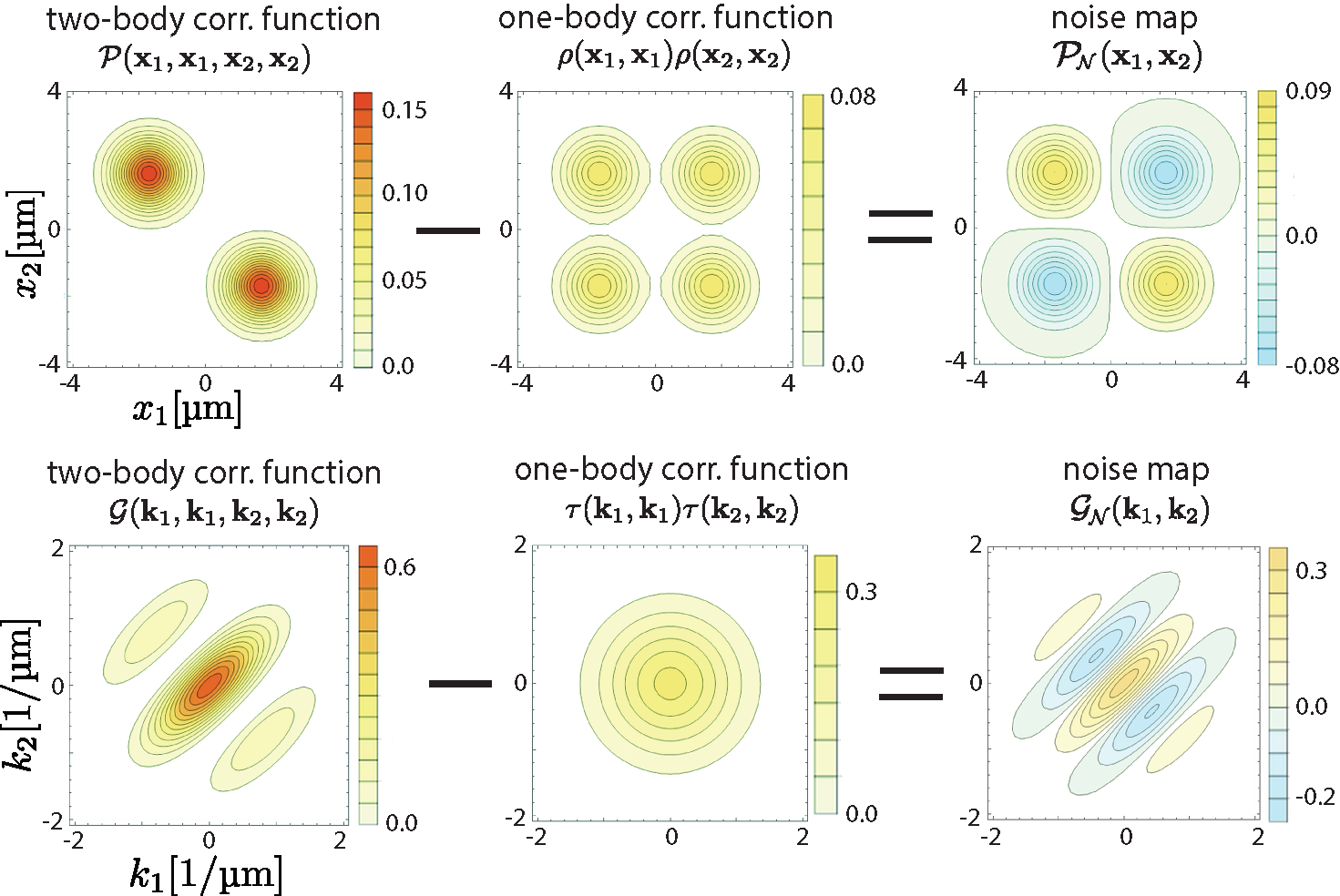} 
\caption{Illustration of the noise-map calculation [see Eqs.\ (\ref{anpn}) and 
(\ref{angn})], for two particles in a single well, separated by a distance $2d={3.4}$ $\mu$m. The noise 
maps are obtained by subtracting the product of one-body correlation functions for the two particles from the 
two-body correlation function. The first row shows the real space correlation functions and the resulting 
noise map after subtraction. The second row shows the same calculation in momentum space. The contour 
levels and the color scheme are the same throughout each row. In the analytic formulas, $s=0.71$ $\mu$m.}
\label{fig:noisemapillustration}  
\end{figure*}

\textcolor{black}{
${\cal C}(s)$ is an overall normalization constant (different at each one of the above formulas) whose precise 
value can be easily determined for a given numerical $s$ value.}
The subtraction process is illustrated in Fig.\ \ref{fig:noisemapillustration} where we plot the real space and 
momentum correlation functions and the corresponding noise map obtained after subtraction. The 
resulting noise maps can be compared to Fig.\ \ref{fig1_2}(q,r), corresponding to the Wigner molecule case. 
The remaining differences in shape between the analytical and the CI noise maps originate from the spatial 
structure of the Wigner molecule, which is more complicated than the two separated Gaussian functions used in 
the analytical modeling. Nonetheless the dominant features are well reproduced. Note that all features in both 
the one-body and two-body correlation-function maps are positive, whereas the noise map contains patterns with 
opposite signs.  

\section{Analytic modeling: Spin-resolved formulas (\fdlu) for three and four particles}
\label{sr34}

Following the derivation illustrated in Sec.\ \ref{anmdtp}, we may generalize it to the cases of $N=3$ and 
$N=4$ particles; see the corresponding effective Heisenberg Hamiltonians given in Appendix 
\ref{sec:hsberg_hamil_and_eigenfunction}. The resulting spin-resolved expressions for the two-body 
correlations of three and four particles can be rather long, but they can be greatly simplified assuming that 
the Gaussian functions are equally spaced and far enough separated so that they have 
negligible overlap.

Here we present results for the ``fixed down look up'' (\fdlu) spin configuration. 

\noindent For three Gaussians centered at $d_1=-2d$, $d_2=0$ and $d_3=2d$ we obtain:

\begin{widetext}

\begin{align}
\begin{split}
\mathcal{P}^{N=3}_{\downarrow\uparrow}({x}_1,{x}_2)=
&{\cal C}(s)\frac{e^{-\frac{8 d^2+6 d (x_1+x_2)+x_1^2+x_2^2}{2 s ^2}}}{36 \pi  s ^2} 
\left(-2 e^{\frac{3 d (x_1+x_2)}{s ^2}}+4 e^{\frac{2 d (d+x_1+x_2)}{s ^2}}+
4 e^{\frac{d (2 d+3 x_1+x_2)}{s ^2}}+\right.\\&\left.e^{\frac{d (5 x_1+x_2)}{s ^2}}+
e^{\frac{d (2 d+x_1+3 x_2)}{s ^2}}+e^{\frac{d (2 d+5 x_1+3 x_2)}{s ^2}}+
e^{\frac{d (x_1+5 x_2)}{s ^2}}+4 e^{\frac{d (2 d+3 x_1+5 x_2)}{s ^2}}+
4 e^{\frac{2 d (d+2 (x_1+x_2))}{s ^2}}\right)
\end{split}
\end{align}

\noindent and

\begin{align}
{\cal G}^{N=3}_{\downarrow\uparrow} (k_1,k_2)=
-{\cal C}(s)\frac{2 s ^2 e^{-2 s ^2 \left(k_1^2+k_2^2\right)}}{9 \pi } 
\bigg( -4 \cos [2 d (k_1-k_2)]+\cos [4 d (k_1-k_2)]-6 \bigg).
\label{eq:spinresolved3particles}
\end{align}

\noindent For four Gaussians centered at $d_1=-3d$, $d_2=-d$, $d_3=d$, and $d_4=3d$ we obtain

\begin{align}
\begin{split}
\mathcal{P}^{N=4}_{\downarrow\uparrow}({x}_1,{x}_2)=
&{\cal C}(s) \frac{e^{-\frac{18 d^2+6 d x_1-6 d x_2+x_1^2+x_2^2}{2 s ^2}}}{144 \pi  s ^2} 
\left(\left(\sqrt{3}-4\right) \left(-e^{\frac{2 d (2 d+x_1)}{s ^2}}\right)+\left(\sqrt{3}+
4\right) e^{\frac{4 d (d+x_1)}{s ^2}}-\right.\\&\left.\left(\sqrt{3}-
4\right) e^{\frac{2 d (2 d+3 x_1-2 x_2)}{s ^2}}-\left(\sqrt{3}-
4\right) e^{\frac{2 d (2 d+2 x_1-3 x_2)}{s ^2}}+
4 e^{\frac{3 d (x_1-x_2)}{s ^2}}+4 e^{\frac{6 d (x_1-x_2)}{s ^2}}+\right.\\
&\left.4 e^{\frac{4 d (2 d+x_1-x_2)}{s ^2}}+4 e^{\frac{2 d (4 d+x_1-x_2)}{s ^2}}-4 \left(\sqrt{3}-
1\right) e^{\frac{2 d (2 d+2 x_1-x_2)}{s ^2}}-4 \left(\sqrt{3}-
1\right) e^{\frac{2 d (2 d+x_1-2 x_2)}{s ^2}}+
\right.\\&\left.4 e^{\frac{d (8 d+3 x_1-3 x_2)}{s ^2}}+
4 \left(\sqrt{3}+1\right) e^{\frac{d (4 d+5 x_1-x_2)}{s ^2}}+
4 \left(\sqrt{3}+1\right) e^{\frac{d (4 d+x_1-5 x_2)}{s ^2}}+
\right.\\&\left.\left(\sqrt{3}+4\right) e^{\frac{2 d (2 d+3 x_1-x_2)}{s ^2}}+
\left(\sqrt{3}+4\right) e^{\frac{2 d (2 d+x_1-3 x_2)}{s ^2}}-
\left(\sqrt{3}-4\right) e^{\frac{2 d (2 d-x_2)}{s ^2}}+
\right.\\&\left.\left(\sqrt{3}+4\right) e^{\frac{4 d (d-x_2)}{s ^2}}+4\right)
\end{split}
\end{align}

\noindent and

\begin{align}
\begin{split}
{\cal G}^{N=4}_{\downarrow\uparrow} (k_1,k_2)=
&{\cal C}(s) \frac{s ^2 e^{-2 s ^2 \left(k_1^2+k_2^2\right)}}{9 \pi } \left(\cos (6 d (k_1-k_2))-
2 \left(\sqrt{3}-1\right) \cos (4 d (k_1-k_2))+\right.\\&\left.
\left(2 \sqrt{3}+3\right) \cos (2 d (k_1-k_2))+12\right).
\label{eq:spinresolved4particles}
\end{split}
\end{align}

\noindent 
\textcolor{black}{
Further support for the shortsightedness of the two-body momentum correlation is found through considerations 
of the analytic expression in Eq.\ (\ref{eq:spinresolved4particles}). From that expression, we find that the 
nearest-neighbor ($2d$-term) contribution is a dominant 72.4\% of the total, compared to 16.4\% and 11.2\% 
contributions from the next-nearest-neighbor ($4d$-term) and next-next-nearest-neighbor ($6d$-term), 
respectively. The resulting interference patterns are illustrated in Fig.\ \ref{fig:fdlu_interference_patterns} 
(neglecting the exponential damping) and in Fig.\ \ref{fig:fdlu_correlation_maps}.
}

\begin{figure*}[t]
\centering
\includegraphics[width=14cm]{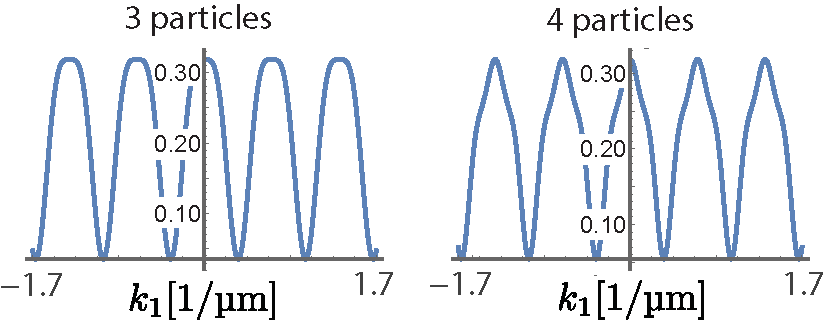}
\caption{Plot of the interference pattern for the fix down look up (\fdlu) spin configuration for three (left) 
and four (right) particles. The interparticle distance (evenly spaced particles) is $2d=4.8$ $\mu$m. The plots 
were obtained by plotting Eq.\ (\ref{eq:spinresolved3particles}) and Eq.\ 
(\ref{eq:spinresolved4particles}) divided by the exponential term 
$e^{-2s^2(k_1^2+k_2^2)}$ and setting $k_2=-k_1$; that is these are cuts along the main cross diagonal 
(top left to bottom right in Fig.\ \ref{fig:fdlu_correlation_maps}). Neither plot shows higher-order 
oscillations since the coefficients of the additional cosine terms are getting increasingly smaller. They 
therefore modify the main oscillation pattern created by the $\cos (2 d (k_1-k_2))$, but do not show additional 
higher frequency oscillations. In the analytic formulas, $s=0.71$ $\mu$m.}
\label{fig:fdlu_interference_patterns}
\end{figure*}

\begin{figure*}[t]
\centering
\includegraphics[width=14cm]{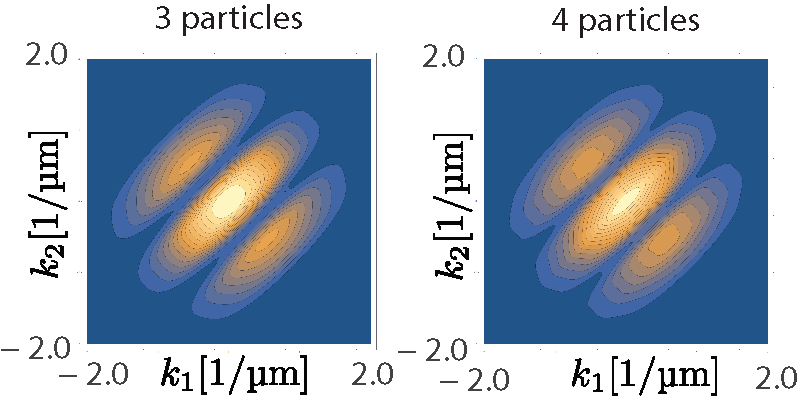}
\caption{Spin-resolved (\fdlu) two-body momentum correlation maps for three (left) and four (right) particles 
at a distance 
$2d=4.8$ $\mu$m. The plots were obtained by plotting Eq.\ (\ref{eq:spinresolved3particles}) (left) and 
Eq.\ (\ref{eq:spinresolved4particles}) (right) with $k_1$ on the x-axis and $k_2$ on the y-axis. In this figure 
further oscillatory pattern beyond the ones shown are damped by the exponential factors.
In the analytic formulas, $s=0.71$ $\mu$m.}
\label{fig:fdlu_correlation_maps}
\end{figure*}

\section{Analytic modeling: Spin unresolved formulas for three and four particles}
\label{su34}

For the following spin-unresolved expressions we restricted ourselves to the same simplifications as in the 
spin-resolved case in the previous section. Namely the Gaussians are equally spaced and far enough apart so 
that their overlap can be neglected.

\noindent For three Gaussians centered at $d_1=-2d$, $d_2=0$, and $d_3=2d$, we obtain:

\begin{align}
\begin{split}
\mathcal{P}^{N=3}({x}_1,{x}_2)=
& {\cal C}(s) \frac{e^{-\frac{8 d^2+6 d (x_1+x_2)+x_1^2+x_2^2}
{2 s ^2}}}{12 \pi s ^2} \left(-2 e^{\frac{3 d (x_1+x_2)}{s ^2}}+
e^{\frac{2 d (d+x_1+x_2)}{s ^2}}+e^{\frac{d (2 d+3 x_1+x_2)}{s ^2}}+
\right.\\&\left.e^{\frac{d (5 x_1+x_2)}{s ^2}}+e^{\frac{d (2 d+x_1+3 x_2)}
{s ^2}}+e^{\frac{d (2 d+5 x_1+3 x_2)}{s ^2}}+e^{\frac{d (x_1+5 x_2)}
{s ^2}}+e^{\frac{d (2 d+3 x_1+5 x_2)}{s ^2}}+e^{\frac{2 d (d+2 (x_1+x_2))}{s ^2}}\right)
\end{split}
\end{align}

\noindent and

\begin{align}
{\cal G}^{N=3} (k_1,k_2)=
{\cal C}(s) \frac{2 s ^2 e^{-2 s ^2 \left(k_1^2+k_2^2\right)}}{3 \pi } 
(\cos (2 d (k_1-k_2))-\cos (4 d (k_1-k_2))+3).
\label{eq:spinunresolved3particles}
\end{align}

\noindent For four Gaussians centered at $d_1=-3d$, $d_2=-d$, $d_3=d$, and $d_4=3d$, we obtain: 

\begin{align}
\begin{split}
\mathcal{P}^{N=4}({x}_1,{x}_2)=
& {\cal C}(s) \frac{e^{-\frac{18 d^2+6 d x_1-6 d x_2+x_1^2+x_2^2}{2 s ^2}}}{24 \pi  s ^2} 
\left(e^{\frac{4 d (d+x_1)}{s ^2}}+e^{\frac{2 d (2 d+x_1)}{s ^2}}+\sqrt{3} 
\left(-e^{\frac{2 d (2 d+2 x_1-x_2)}{s ^2}}\right)+\right.\\
&\left.\sqrt{3} e^{\frac{d (4 d+5 x_1-x_2)}{s ^2}}-\sqrt{3} e^{\frac{2 d (2 d+x_1-2 x_2)}{s ^2}}+
\sqrt{3} e^{\frac{d (4 d+x_1-5 x_2)}{s ^2}}+e^{\frac{6 d (x_1-x_2)}{s ^2}}+\right.\\
&\left.e^{\frac{4 d (2 d+x_1-x_2)}{s ^2}}+e^{\frac{2 d (4 d+x_1-x_2)}{s ^2}}+
e^{\frac{2 d (2 d+3 x_1-x_2)}{s ^2}}+e^{\frac{2 d (2 d+3 x_1-2 x_2)}{s ^2}}+
e^{\frac{2 d (2 d+x_1-3 x_2)}{s ^2}}+\right.\\
&\left.e^{\frac{2 d (2 d+2 x_1-3 x_2)}{s ^2}}+e^{\frac{4 d (d-x_2)}{s ^2}}+
e^{\frac{2 d (2 d-x_2)}{s ^2}}+1\right)
\end{split}
\end{align}

\noindent and

\begin{align}
{\cal G}^{N=4} (k_1,k_2)= 
{\cal C}(s) \frac{s ^2 e^{-2 s ^2 \left(k_1^2+k_2^2\right)}}{3 \pi } \left(\sqrt{3} 
\cos (2 d (k_1-k_2))-\sqrt{3} \cos (4 d (k_1-k_2))+6\right).
\label{eq:spinunresolved4particles}
\end{align}

\noindent The resulting interference patterns are plotted in Fig.\ 
\ref{fig:spin_unresolved_interference_patterns} (neglecting the exponential damping) and 
Fig.\ \ref{fig:spin_unresolved_correlation_maps}.

\begin{figure*}[t]
\centering
\includegraphics[width=14cm]{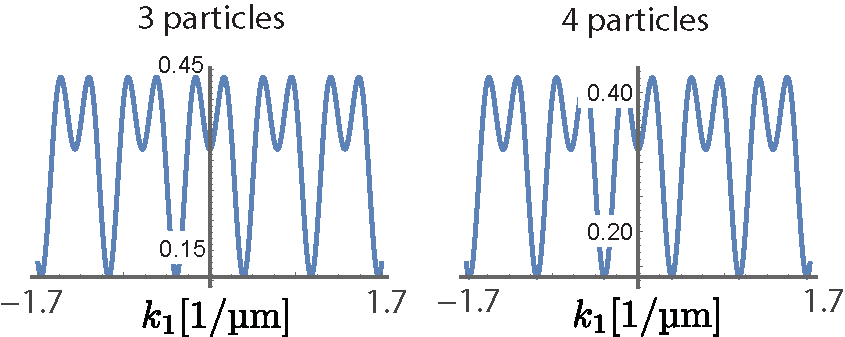}
\caption{Plot of the interference pattern in the spin unresolved case for three (left) and four (right) 
particles. The interparticle distance (evenly spaced particles) is $2d=4.8$ $\mu$m. The plots were obtained by 
plotting Eq.\ (\ref{eq:spinunresolved3particles}) and Eq.\ (\ref{eq:spinunresolved4particles}) divided by the 
exponential term $e^{-2s^2(k_1^2+k_2^2)}$. It is remarkable that the plots for three and four particles are 
very similar and their difference is only the overall scaling ($\sqrt{3}/2$) and a constant shift 
($(6-3\sqrt{3})/(6\pi)$). In the analytic formulas, $s=0.71$ $\mu$m.}
\label{fig:spin_unresolved_interference_patterns}
\end{figure*}

\begin{figure*}[t]
\centering
\includegraphics[width=14cm]{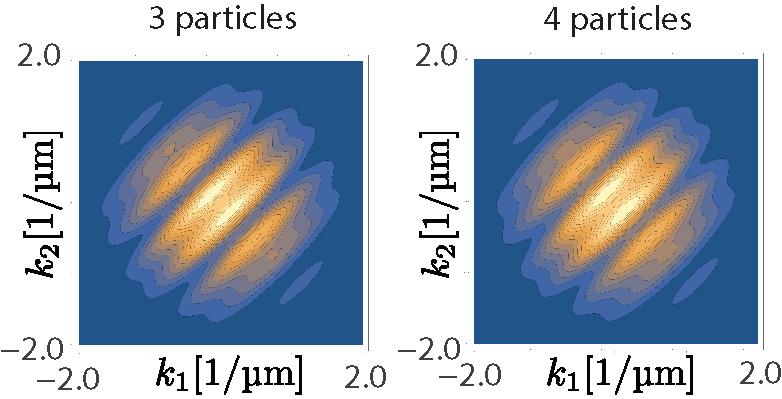}
\caption{Spin unresolved correlation maps for three (left) and four (right) particles at a distance 
$2d=4.8$ $\mu$m (evenly spaced particles). The plots were obtained by plotting Eq.\ 
(\ref{eq:spinunresolved3particles}) (left) and Eq.\ (\ref{eq:spinunresolved4particles}) (right) with $k_1$ on 
the $x$-axis and $k_2$ on the $y$-axis. In the analytic formulas, $s=0.71$ $\mu$m.}
\label{fig:spin_unresolved_correlation_maps}
\end{figure*}

\textcolor{black}{
\section{Effective Heisenberg Hamiltonians and corresponding ground state solutions for three and four particles}
\label{sec:hsberg_hamil_and_eigenfunction}
}

\textcolor{black}{
Here we give for the readers' convenience the effective Heisenberg Hamiltonians for three and four particles and
their corresponding ground state eigenvectors. We note again that, for a small number of repelling trapped 
particles (electrons in semiconductor quantum dots and ultracold fermions or bosons), the mapping of the 
microscopic many-body Hamiltonian onto spin-chain-type, effective Heisenberg Hamiltonians has been demonstrated 
recently and it constitutes an ongoing active area of research; for electrons in semiconductor quantum dots
see Refs.\ \cite{li07,li09}, for  ultracold fermions or bosons in quasi-1D traps see Refs.\
\cite{yann15,yann16,zinn14,deur14,zinn15,bruu15,mass15,murm15,murm15.2,pu15,cui16}.}

The three particle Heisenberg Hamiltonian in matrix form with spin primitives 
$|\uparrow\uparrow\downarrow\rangle,|\uparrow\downarrow\uparrow\rangle,|\downarrow\uparrow\uparrow\rangle$ 
is given as 

\begin{align}
H=  \begin{pmatrix}
  0 & J & 0 \\
  J & -J & J \\
  0 & J & 0
\end{pmatrix},
\end{align}

\noindent the corresponding ground state eigenfunction is 

\begin{align}
v_1=\frac{1}{\sqrt{6}}|\uparrow\uparrow\downarrow\rangle-
\sqrt{\frac{2}{3}}|\uparrow\downarrow\uparrow\rangle+\frac{1}{\sqrt{6}}|\downarrow\uparrow\uparrow\rangle.
\end{align}

\noindent For four particles the Hamiltonian in matrix form with spin primitives 
$|\uparrow\uparrow\downarrow\downarrow\rangle,|\uparrow\downarrow\uparrow\downarrow\rangle,
|\downarrow\uparrow\uparrow\downarrow\rangle,|\uparrow\downarrow\downarrow\uparrow\rangle,
|\downarrow\uparrow\downarrow\uparrow\rangle,|\downarrow\downarrow\uparrow\uparrow\rangle$ is given as

\begin{align}
H=\begin{pmatrix}
-\frac{J_{23}}{4}+\frac{J_{12}}{2} & \frac{J_{23}}{2} & 0 & 0 & 0 & 0\\
\frac{J_{23}}{2} & -\frac{J_{23}}{4}-\frac{J_{12}}{2} & \frac{J_{12}}{2} & \frac{J_{12}}{2} & 0 & 0\\
0 & \frac{J_{12}}{2} & \frac{J_{23}}{4}-\frac{J_{12}}{2} & 0 & \frac{J_{12}}{2} & 0\\
0 & \frac{J_{12}}{2} & 0 & \frac{J_{23}}{4}-\frac{J_{12}}{2} & \frac{J_{12}}{2} & 0\\
0 & 0 & \frac{J_{12}}{2} & \frac{J_{12}}{2} & -\frac{J_{23}}{4}-\frac{J_{12}}{2} & \frac{J_{23}}{2}\\
0 & 0 & 0 & 0 & \frac{J_{23}}{2} & -\frac{J_{23}}{4}+\frac{J_{12}}{2}
\end{pmatrix},
\end{align}

\noindent where we have set $J_{34}=J_{12}$ due to symmetry (we have equally spaced wells). For well-separated
wells, one can further approximate $J_{12} \approx J_{23}$. Then the corresponding ground-state eigenvector is

\begin{equation}    
    \begin{split}
v_1=\frac{1}{\sqrt{2+2(1+\sqrt{3})^2+2(2+\sqrt{3})^2}} \left(
|\uparrow\uparrow\downarrow\downarrow\rangle-(2+\sqrt{3})
|\uparrow\downarrow\uparrow\downarrow\rangle+(1+\sqrt{3})
|\downarrow\uparrow\uparrow\downarrow\rangle+ \right. \\ 
\left.  (1+\sqrt{3})|\uparrow\downarrow\downarrow\uparrow\rangle-(2+\sqrt{3})
|\downarrow\uparrow\downarrow\uparrow\rangle+|\downarrow\downarrow\uparrow\uparrow\rangle \right).
    \end{split} 
    \end{equation}
\end{widetext}

\end{document}